\documentclass[sigconf,screen]{acmart}
\usepackage{subcaption}
\usepackage[normalem]{ulem}
\usepackage{booktabs}
\usepackage[nounderscore]{syntax}
\usepackage{xspace}
\usepackage{xparse}
\usepackage{microtype}
\usepackage{listings, multicol}
\usepackage{tikz}
\usepackage{colortbl}
\usepackage{float}
\usepackage{amsbsy}
\usepackage{wrapfig}
\usepackage{mathtools}
\usepackage{enumerate}
\usepackage{enumitem}
\usepackage{varwidth}
\usepackage{multicol}
\usepackage{multirow}
\usepackage{tabularx}
\usepackage{outlines}
\usepackage[toc,titletoc,page,title]{appendix}
\usepackage{amsmath}
\usepackage{pgfplots}
\usepackage{pgfplotstable}
\usepackage{xcolor,colortbl,xcolor-solarized}
\usepackage{amsthm}
\usepackage{cleveref}
\usepackage{tabu}
\usepackage{mfirstuc}
\usepackage{array}
\usepackage{outlines}
\usepackage{amsfonts}
\usepackage{pifont}
\usepackage{tablefootnote}
\usepackage[frozencache,cachedir=.]{minted}
\pgfplotsset{compat=1.7}

\usepackage[bottom]{footmisc}

\usepackage{xargs}                  
\usetikzlibrary{positioning, calc, patterns, shapes.misc, fit, backgrounds, matrix}

\makeatletter
\def\th@definition{%
  \thm@notefont{}
}
\makeatother

\theoremstyle{plain}
\newtheorem{theorem}{Theorem}[section]

\theoremstyle{definition}

\newtheorem{block}[theorem]{Definition}

\newcommand{\figref}[1]{\Cref{#1}}

\newcommand{\alg}[1]{algorithm#1}
\newcommand{\crdtree}[1]{fibertree#1}
\newcommand{\Crdtree}[1]{Fibertree#1}
\newcommand{\rdscan}[1]{level scanner#1}
\newcommand{\Rdscan}[1]{Level scanner#1}
\newcommand{\wrscan}[1]{level writer#1}

\newcommand{\repeater}[1]{repeater#1}
\newcommand{\fiber}[1]{fiber#1}

\newcommand{\ntkn}[1]{empty#1}
\newcommand{\numerical}[1]{non-control#1}

\newcommand{\hide}[1]{}

\newcommand{\sam}{SAM}
\newcommand{\samlong}{Sparse Abstract Machine}
\newcommand{\samCoreBlockCount}{nine}

\newcommand{\spmspm}{SpM*SpM}

\newcommand{\dataflow}{dataflow}

\AtBeginDocument{%
  \providecommand\BibTeX{{%
    \normalfont B\kern-0.5em{\scshape i\kern-0.25em b}\kern-0.8em\TeX}}}

\setcopyright{rightsretained}
\acmPrice{}
\acmDOI{10.1145/3582016.3582051}
\acmYear{2023}
\copyrightyear{2023}
\acmSubmissionID{asplosc23main-p336-p}
\acmISBN{978-1-4503-9918-0/23/03}
\acmConference[ASPLOS '23]{Proceedings of the 28th ACM International Conference on Architectural Support for Programming Languages and Operating Systems, Volume 3}{March 25--29, 2023}{Vancouver, BC, Canada}
\acmBooktitle{Proceedings of the 28th ACM International Conference on Architectural Support for Programming Languages and Operating Systems, Volume 3 (ASPLOS '23), March 25--29, 2023, Vancouver, BC, Canada}
\received{2022-10-20}
\received[accepted]{2023-01-19}

\begin{document}

\title{The Sparse Abstract Machine}

\author{Olivia Hsu}
\orcid{0000-0002-4195-8106}
\affiliation{%
  \institution{Stanford University}
  % \streetaddress{P.O. Box 1212}
    \country{USA}
}
\email{owhsu@stanford.edu}

\author{Maxwell Strange}
\orcid{0000-0001-5945-1349}
\affiliation{%
  \institution{Stanford University}
  % \streetaddress{P.O. Box 1212}
    \country{USA}
}
\email{mstrange@stanford.edu}

\author{Ritvik Sharma}
\orcid{0000-0002-5809-7031}
\affiliation{%
  \institution{Stanford University}
  % \streetaddress{P.O. Box 1212}
    \country{USA}
}
\email{rsharma3@stanford.edu}

\author{Jaeyeon Won}
\orcid{0000-0002-3082-4348}
\affiliation{%
  \institution{MIT}
  \country{USA}
}
\email{jaeyeon@mit.edu}

\author{Kunle Olukotun}
\orcid{0000-0002-8779-0636}
\affiliation{%
  \institution{Stanford University}
  % \streetaddress{P.O. Box 1212}
    \country{USA}
}
\email{kunle@stanford.edu}

\author{Joel S. Emer}
\orcid{0000-0002-3459-5466}
\affiliation{%
  \institution{MIT and NVIDIA}
  \country{USA}
}
%\additionalaffiliation{%
%  \institution{NVIDIA}
%  \country{USA}
%}
\email{jsemer@mit.edu}

\author{Mark A. Horowitz}
\orcid{0000-0003-3245-7542}
\affiliation{%
  \institution{Stanford University}
  % \streetaddress{P.O. Box 1212}
    \country{USA}
}
\email{horowitz@ee.stanford.edu}

\author{Fredrik Kj\o lstad}
\orcid{0000-0002-2267-903X}
\affiliation{%
  \institution{Stanford University}
  % \streetaddress{P.O. Box 1212}
    \country{USA}
}
\email{kjolstad@stanford.edu}

\renewcommand{\shortauthors}{O. Hsu, M. Strange, R. Sharma, J. Won, K. Olukotun, J. S. Emer, M. A. Horowitz, and F. Kj{\o}lstad}

\begin{abstract}
We propose the Sparse Abstract Machine (SAM), an abstract machine model for targeting sparse tensor algebra to reconfigurable and fixed-function spatial dataflow accelerators. SAM defines a streaming dataflow abstraction with sparse primitives that encompass a large space of scheduled tensor algebra expressions. SAM dataflow graphs naturally separate tensor formats from algorithms and are expressive enough to incorporate arbitrary iteration orderings and many hardware-specific optimizations. We also present Custard, a compiler from a high-level language to SAM that demonstrates SAM's usefulness as an intermediate representation. We automatically bind from SAM to a streaming dataflow simulator. We evaluate the generality and extensibility of SAM, explore the performance space of sparse tensor algebra optimizations using SAM, and show SAM's ability to represent dataflow hardware. 
\end{abstract}





\maketitle

\section{Introduction}

Specialized streaming \dataflow{} accelerators that leverage pipelining, locality, and parallelism are becoming increasingly popular as performance- and energy-efficient alternatives to CPUs and GPUs. 
But the efficiency comes at the cost of programmability: all have limits to their application domain, and most have limited and/or difficult programming interfaces. As a result, users often access these accelerators through library calls that are created by expert programmers~\cite{vilim2021aurochs,vilim2020gorgon}.  
Recent research has proposed closing this flexibility and programmability gap by creating reconfigurable dataflow architectures or coarse-grained reconfigurable arrays~\cite{plasticine,nowatzki2017, spu, fifer2021nguyen,carsello2022amber, hegde2019extensor, triginst2013parashar}, including compilation tools to map a class of user applications to these arrays~\cite{zhang2021sara, spatial, liu2021unifiedbuffer, pu2017, weng2022unifying}.

Given these trends, it is not surprising that interest in general accelerators for sparse tensor algebra is increasing~\cite{hegde2019extensor, spu, rucker2021capstan}.
Sparse tensor algebra has applications across many fields including science, engineering, data and graph analytics, and machine learning~\cite{feynman_physics, moleculardynamics_sim, kolda2008datamining, kepner2011graphalgs, abadi2016tensorflow, scnn}. Tensor algebra generalizes linear algebra to higher-order tensors, and ``sparse'' indicates tensor algebra computations where one or more tensors are stored in compressed data structures that omit zeros. Sparse tensor algebra, expressed as a language using tensor index notation or Einstein summation (Einsum) notation, is an important language with a long history, starting as a mathematical notation~\cite{Ricci1901}. It has recently gained traction as a computational language~\cite{BaKo07ttb} that subsumes linear algebra. To accelerate these computations, many papers have also been published on point solutions for single-expression hardware, where the expression is often sparse matrix multiplication~\cite{pal2018outerspace,qin2021sigma,zhang2020sparch,zhang2021gamma,tensaurus,srivastava2020matraptor}.

Most sparse tensor algebra accelerators are fixed-function matrix multiply engines. Arbitrary sparse tensor contractions must therefore be reduced to sparse matrix multiplications through algebraic \textit{factorization}~\cite{ctf-sparse, ctf-main}. Factorization breaks up large expressions using transpositions, tensor-to-matrix conversions, and temporaries.
However, compared to dense tensor algebra, factorization is significantly more expensive for sparse tensor algebra. In fact, a sequence of matrix multiplications is often more than an order of magnitude slower compared to bespoke generated tensor contractions. 
And, more importantly, the lack of fusion in sparse computations can, and often does, lead to inferior worst-case asymptotic complexity: the runtime of an unfused expression may grow with the number of tensor components while a fused expression grows with the number of nonzero components~\cite{kjolstad2020sparse}.
To address the limitations of fixed-function accelerators, the SPU~\cite{spu}, ExTensor~\cite{hegde2019extensor}, and Capstan~\cite{rucker2021capstan} systems propose programmable sparse dataflow hardware. However, they lack full support for sparse tensor algebra.

We define an abstract machine model for sparse tensor algebra called the Sparse Abstract Machine (\sam{}) that accelerates general sparse tensor algebra. 
\sam{} consists of abstract dataflow blocks that lend themselves to VLSI implementations and compose to implement any sparse tensor algebra expression and to implement many algorithms for each expression, including fused algorithms, unfused algorithms with temporaries, tiled algorithms, and parallelized and vectorized algorithms.
Thus, \sam{} can simultaneously be used to analyze point solutions, be the abstract architecture of a programmable sparse tensor algebra \dataflow{} implementation, and be the intermediate representation of its compiler.

We built \sam{} to be for sparse tensor dataflow accelerators what LLVM~\cite{lattner04llvm} is for instruction-based conventional processors: It defines the machine functionality and provides an interface between the compiler and hardware, allowing the end-to-end system to continue to function while both sides are independently optimized. \sam{} also enumerates the primitives that are needed to support all features of sparse tensor algebra.
Our contributions are: 
\begin{enumerate}
    \item the first abstract machine model that expresses the whole of sparse tensor algebra computations as spatial dataflow graphs on multidimensional sparse and dense tensors,
    \item cleanly defined dataflow primitives for each of the fundamental features of sparse tensor algebra,
    \item a representation of multidimensional sparse/dense tensors as flattened streams with hierarchical control tokens, and
    \item a compilation strategy from a high-level tensor index notation to our abstract machine model.
\end{enumerate}
To evaluate our contributions we implemented \sam{} as a cycle-approximate simulator that is generated by our compiler, Custard. Using the simulations, we search the space of sparse tensor algebra architectural designs. Finally, we show that SAM can represent prior sparse dataflow accelerators.

\section{Background}
\label{sec:background}

This section describes the necessary features of a general sparse tensor algebra computing system. We discuss how these features can be programmed with the input APIs of the TACO compiler~\cite{kjolstad2017tensor}. We describe why TACO only targets von Neumann machines and propose a new compiler in \Cref{sec:compiler} that uses SAM to target dataflow accelerators. Finally, we discuss limitations of prior work on fixed-function and programmable sparse tensor algebra hardware.

\subsection{The Design Space of Sparse Tensor Algebra}
\label{sec:features}

Tensor algebra computations are typically expressed using tensor index notation (or Einsum notation), where tensors are indexed by index variables, are multiplied and added, and where results may be summed over index variables. For example, matrix multiplication can be written in tensor index notation as $X_{ij} = \sum_k B_{ik} C_{kj}$, where index variables $i$, $j$, and $k$ range over the rows and columns that they index. Expressions may have more than two operands, such as sampled dense-dense matrix multiplication (SDDMM) $X_{ij} = \sum_k B_{ij}C_{ik}D_{jk}$. For such compound expressions, it is often beneficial to fuse the resulting computation (i.e., loop fusion or hardware pipelining to avoid materializing large temporary data structures). Tensor index notation only specifies the expression (or algorithm of computation) and does not does not include a description of the schedule (e.g. dataflow traversal order, tiling methodology, and parallelization). Prior work popularizes the separation of algorithm and schedule in both software~\cite{halide2012, venkat2015chillie, senanayake2020, tvm2018} and hardware~\cite{eyeriss2017chen}. 

Tensor index notation consists of five features that must be supported by any general tensor algebra computing system:
\begin{enumerate}
    \item a way to traverse multidimensional tensors;
    \item a way to combine traversal over multiple tensors;
    \item a way to repeat operands over other operands, e.g., in $x_i=\sum_j B_{ij} c_j,$ $c$ must be multiplied by each row of $B$;
    \item a way to compute scalar additions and multiplications, including summation reductions; and
    \item a way to assign results to a tensor.
\end{enumerate}
Efficient sparse tensor algebra computing systems must also support
\begin{enumerate}
\setcounter{enumi}{5}
    \item compressed data structures for sparse tensors,
    \item index variable iteration in any order, and
    \item fusion of the computation
\end{enumerate}
in order to avoid inferior worst-case asymptotic complexity~\cite{kjolstad2020sparse,ahrens2022}.

\subsection{The TACO Compiler}

The TACO compiler~\cite{kjolstad2017tensor} and related systems~\cite{mutlu2020comet,bik2022compiler,yadav2022, yadav2022spdistal, sparsetir} compile tensor index notation to von Neumann architectures, including CPUs, GPUs~\cite{senanayake2020}, and distributed machines~\cite{yadav2022, yadav2022spdistal}. TACO supports all five features of tensor index notation, as well as compressed data structures~\cite{kjolstad2017tensor,chou2018}, iteration reordering~\cite{kjolstad2019}, and fusion~\cite{kjolstad2017tensor}. 
The TACO compiler has three separate languages that independently describe functionality, data, and optimization: tensor index notation, a data representation language, and a scheduling language.

Although TACO has the generality we described in \Cref{sec:features}, it only compiles to von Neumann architectures. This limitation is due to its lowering machinery fundamentally embedding the \\
(co-)iteration over one or more tensors into general-purpose control flow found in von Neumann machines including: indirect index accesses, \texttt{while} loops, and \texttt{if} statements. The heavy reliance on these constructs during lowering and code generation for traversing irregular structures makes it inapplicable for reconfigurable dataflow accelerators since many of these architectures remove general control flow to achieve higher performance~\cite{plasticine, rucker2021capstan, triginst2013parashar, carsello2022amber, hegde2019extensor, spu}. Converting the complex control flow generated by TACO into a dataflow abstraction, such as the one we describe in this paper, would require significant assumptions about data accesses and/or a complex synthesis system. Instead, we describe an alternative compiler lowering algorithm that lowers from TACO's high-level concrete index notation~\cite{kjolstad2019} to the SAM dataflow abstraction.

\subsection{Prior Work on Fixed-Function Hardware}
\label{sec:background-fixed-hw}

There is a large body of recent work on architectures that explore point solutions in the space of hardened sparse tensor algebra expressions and algorithms~\cite{tensaurus,scnn, han2016eie, pal2018outerspace, srivastava2020matraptor, he2020sparsetpu, chen2018eyerissv2,qin2021sigma,zhang2021gamma}, which we will call fixed-function accelerators. 
Many of these implement sparse matrix multiplication (\spmspm{})~\cite{han2016eie, he2020sparsetpu, qin2021sigma, pal2018outerspace, srivastava2020matraptor, zhang2021gamma}. For example, SIGMA~\cite{qin2021sigma} implements the inner-product $i \rightarrow j \rightarrow k$ iteration order. Although this order is typically preferred for dense matrix multiplications, and although it avoids scattering into the result, it has poor asymptotic performance because it iterates over all combinations of $i$ and $j$ before coordinates are intersected at the contracted variable $k$. GAMMA~\cite{zhang2021gamma} applies Gustavson's~\cite{gustavson1978} $i \rightarrow k \rightarrow j$ iteration order, which improves the asymptotic complexity at the cost of merge hardware to rearrange the reduction step to allow in-order generation of elements of $X$. And OuterSPACE~\cite{pal2018outerspace} implements the outer-product $k \rightarrow i \rightarrow j$ iteration order, which for doubly-compressed sparse row (DCSR) matrices has better asymptotic complexity than the $i \rightarrow k \rightarrow j$ order but requires an additional step for merging whole matrices into $X$.

Using fixed-function matrix multiplication hardware to compute any expression in tensor algebra relies on factorizing general expressions into a sequence of invocations of matrix multiplication operations. Factorizing unfuses the computation and fixes the dataflow ordering. 
\sam{} removes this limitation by having sufficient power to express fused expressions on accelerators, letting us explore both the benefits and the costs of these approaches.

\subsection{Prior Work on Programmable Hardware}
\label{sec:background-reconfig-hw}

The limitations of fixed-function dataflow hardware
motivate programmable sparse tensor algebra dataflow hardware, which are implementations of an abstract machine like the one in this paper. We describe three major lines of prior work on programmable sparse tensor algebra dataflow hardware: the SPU~\cite{spu}, ExTensor~\cite{hegde2019extensor}, and Capstan~\cite{rucker2021capstan}. 
While these designs do not support the full generality described in \Cref{sec:features}, they suggest the essential hardware structures and techniques in dataflow accelerators for sparse tensor algebra and greatly influenced our work.

\subsubsection*{SPU}
The Sparse Processing Unit (SPU)~\cite{nowatzki2017,spu} is a spatial streaming dataflow architecture where instructions on a CPU configure a Coarse-Grained Reconfigurable Architecture (CGRA) and then stream arrays to it. It has efficient hardware both for combining streams (e.g., an intersection) and for using one stream to index into an on-chip array~\cite{spu}. The SPU can be used to implement binary vector operations, relational joins, and graph algorithms~\cite{dadu2021polygraph}, thus unifying domains. The SPU also includes an idiom-directed compiler~\cite{weng2022unifying} from pragma-annotated C loops to CGRA configurations.

Although the SPU literature describes operations that support a broad set of domains---tensor algebra, relational algebra, and graph operations---the SPU CGRA~\cite{spu} only supports vector operations, while higher-order tensor algebra operations appear to be implemented as CPU loops that dispatch inner-loop vector operations to the CGRA (see Figure~6 from Nowatzki et al.~\cite{nowatzki2017}, which the SPU extends). Thus, higher-order expressions must be broken into pieces with data flowing between the CPU and CGRA engine, reducing pipeline locality. 
The abstract machine we describe in this paper provides a complete dataflow model for tensor algebra. Using it, together with the compiler, to target the SPU and the TACO compiler to target the CPU, provides a fruitful path towards compiling to the SPU system from a higher-level tensor algebra language.

\subsubsection*{ExTensor}
The ExTensor system~\cite{hegde2019extensor} is a CGRA-style architecture designed to hierarchically evaluate Einsum operations on sparse tensors. In ExTensor, a compute element is made up of memory that stores pieces of tensors, i.e, fibers, and hardware to perform operations on those fibers (e.g., read/stream them in/out, perform intersections/MACCs). ExTensor's design primarily considered a topology of compute elements that each operated on two fibers (from two operand tensors) at a time, and was focused on computing \spmspm{}. To support SDDMM, which has three operands, an ExTensor instance with additional compute elements was proposed. However, it cannot perform all Einsum computations, such as those with union merges for addition. Extensor also did not provide a programmatic approach to map an arbitrary Einsum to a concrete ExTensor instance, which makes it hard to add new expressions that were not tested by the authors. Nor is there a discussion of how the architecture might change to better suit the needs of arbitrary Einsums. \sam{} and our compiler addresses these limitations, and should allow for the creation of ExTensor configurations.

\subsubsection*{Capstan}
Finally, the Capstan system~\cite{rucker2021capstan} supports hierarchical iteration over dense, compressed, bitvector, and bit-tree data structures. For some data structure types, it supports any tensor algebra expression, including additions and multiplications. Its primary limitation is that it does not support combining two or more compressed data structures (e.g., by intersection) at one iteration level. Instead, it relies on bitvectors and bit-trees, which densify the iteration. The paper argues that this works well for common clustered sparse tensors, but not for arbitrary sparsity. Finally, the Capstan system does not support efficient broadcasting of a tensor over a sparse inner dimension of another tensor, as it relies on programmed counters to control broadcasting.  While Capstan is programmed by Spatial~\cite{spatial, zhang2021sara}, a domain specific language for hardware accelerators based on parallel patterns~\cite{prabhakar2016parallelpatterns}, Spatial is at a lower-level of abstraction---more descriptive of the hardware accelerator---than the Custard input APIs.  Again \sam{} provides an opportunity to provide a higher-level programming interface for this system.

\begin{figure*}[htb]
    \newcommand{\datamodelscale}{0.151}
    \subfloat[Matrix $B_{ij}$]{
        \includegraphics[scale=\datamodelscale]{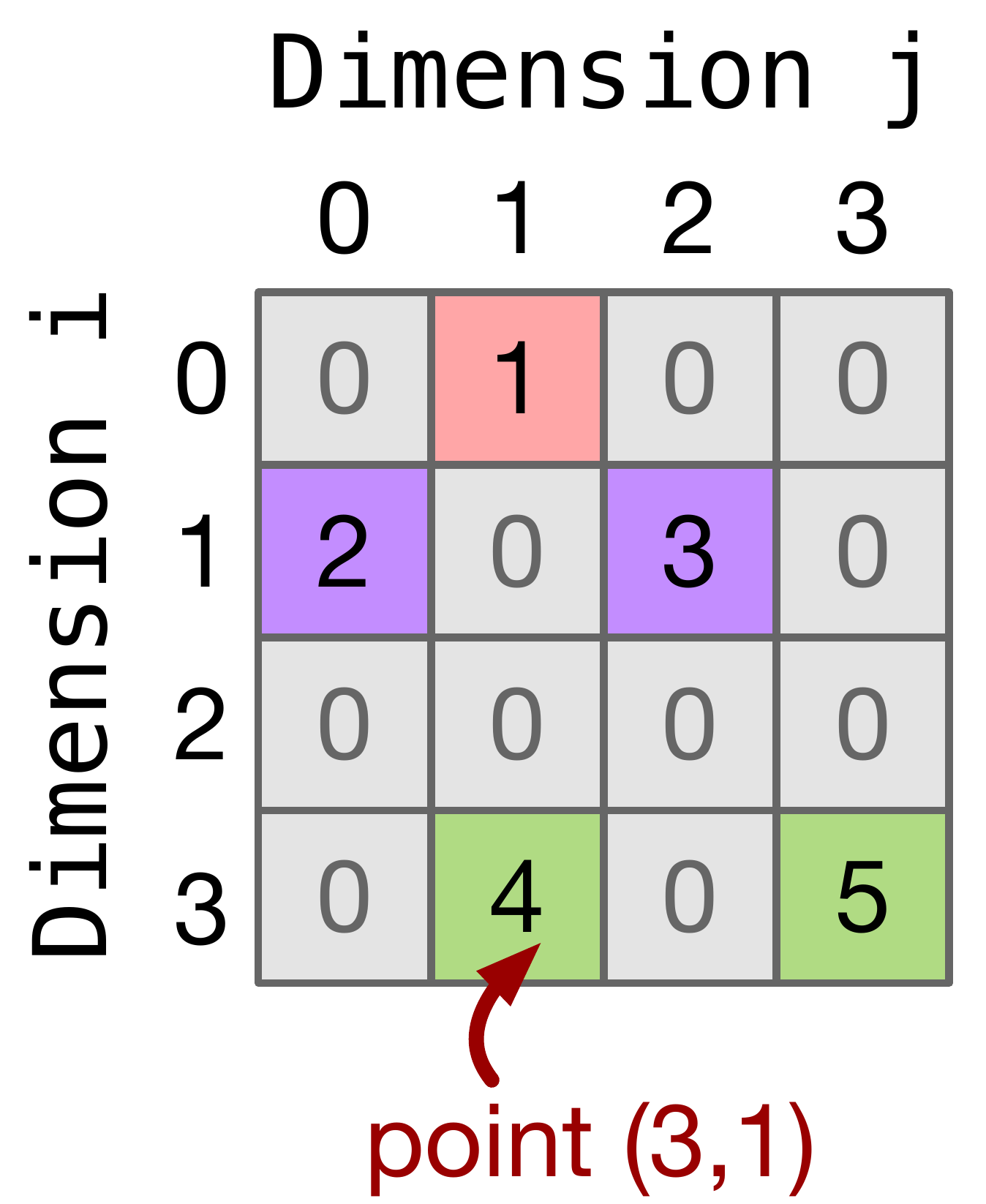}
        \label{fig:data-model-matrix}
    }
    \hfill
    \subfloat[$B_{ij}$ as a \crdtree{}]{
        \includegraphics[scale=\datamodelscale]{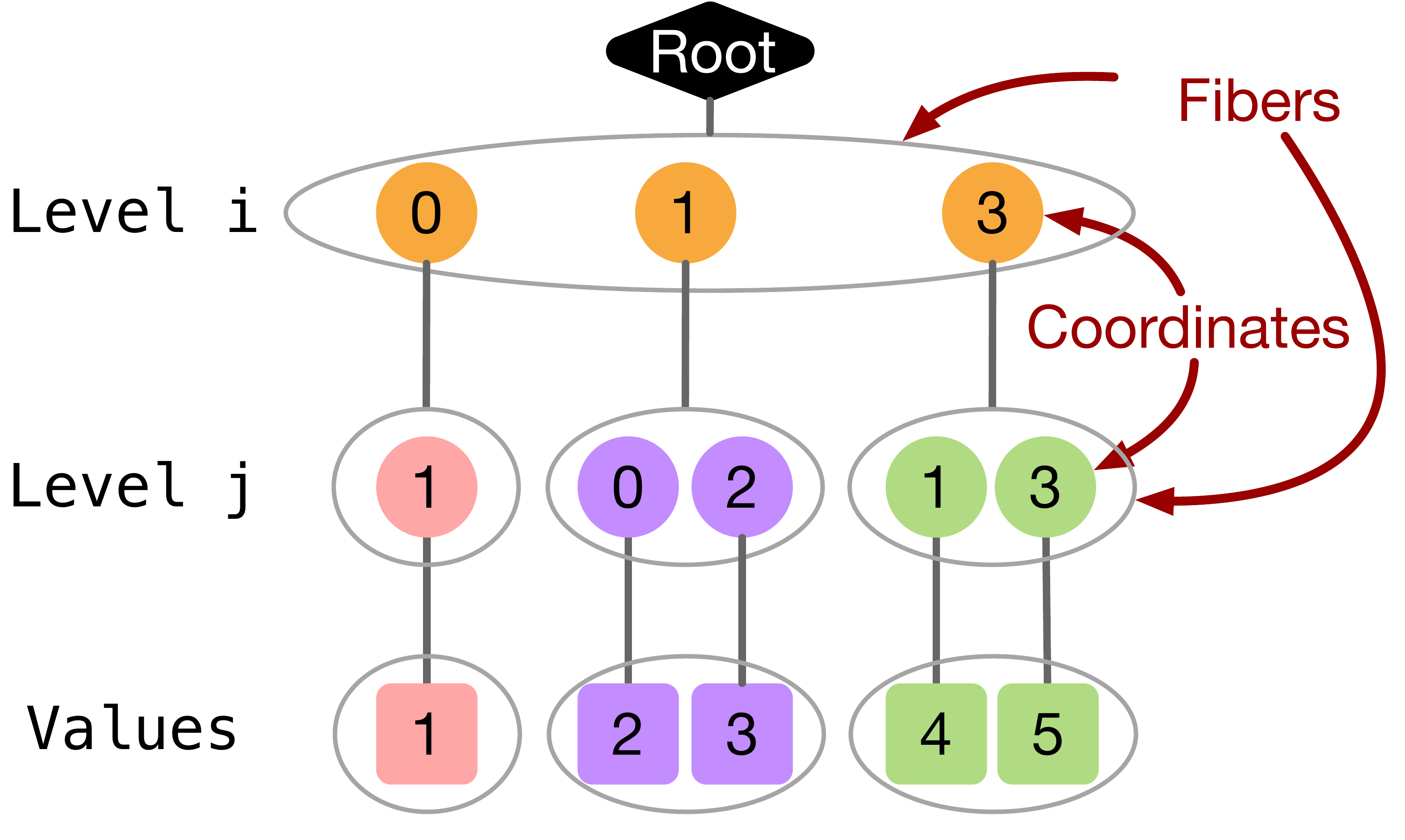}
        \label{fig:data-model-coordinate-tree}
    }
    \hfill
    \subfloat[$B_{ij}$ stored in memory]{
        \includegraphics[scale=\datamodelscale]{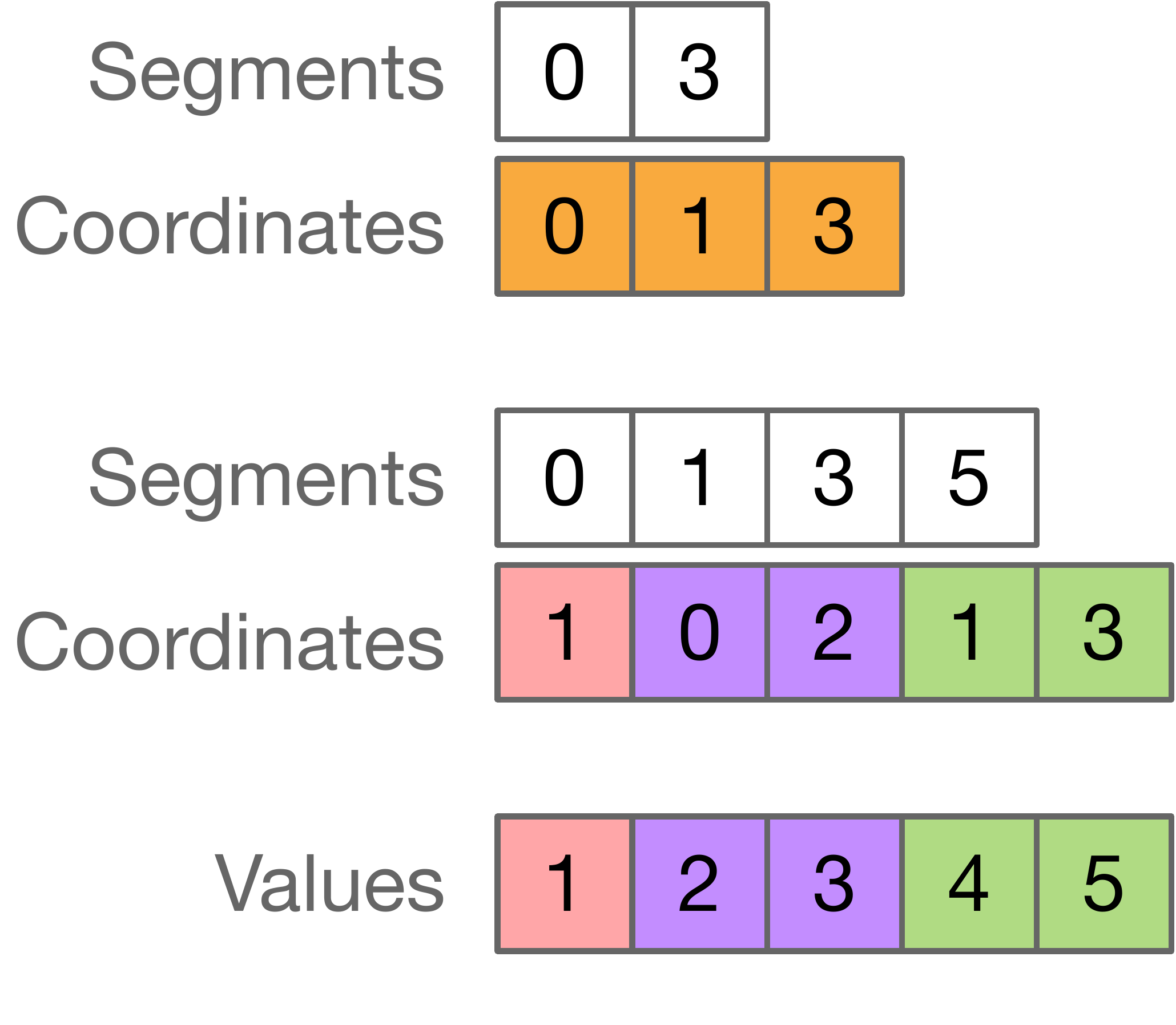}
        \label{fig:data-model-memory}
    }
    \hfill
    \subfloat[$B_{ij}$ sent through a stream]{
        \includegraphics[scale=\datamodelscale]{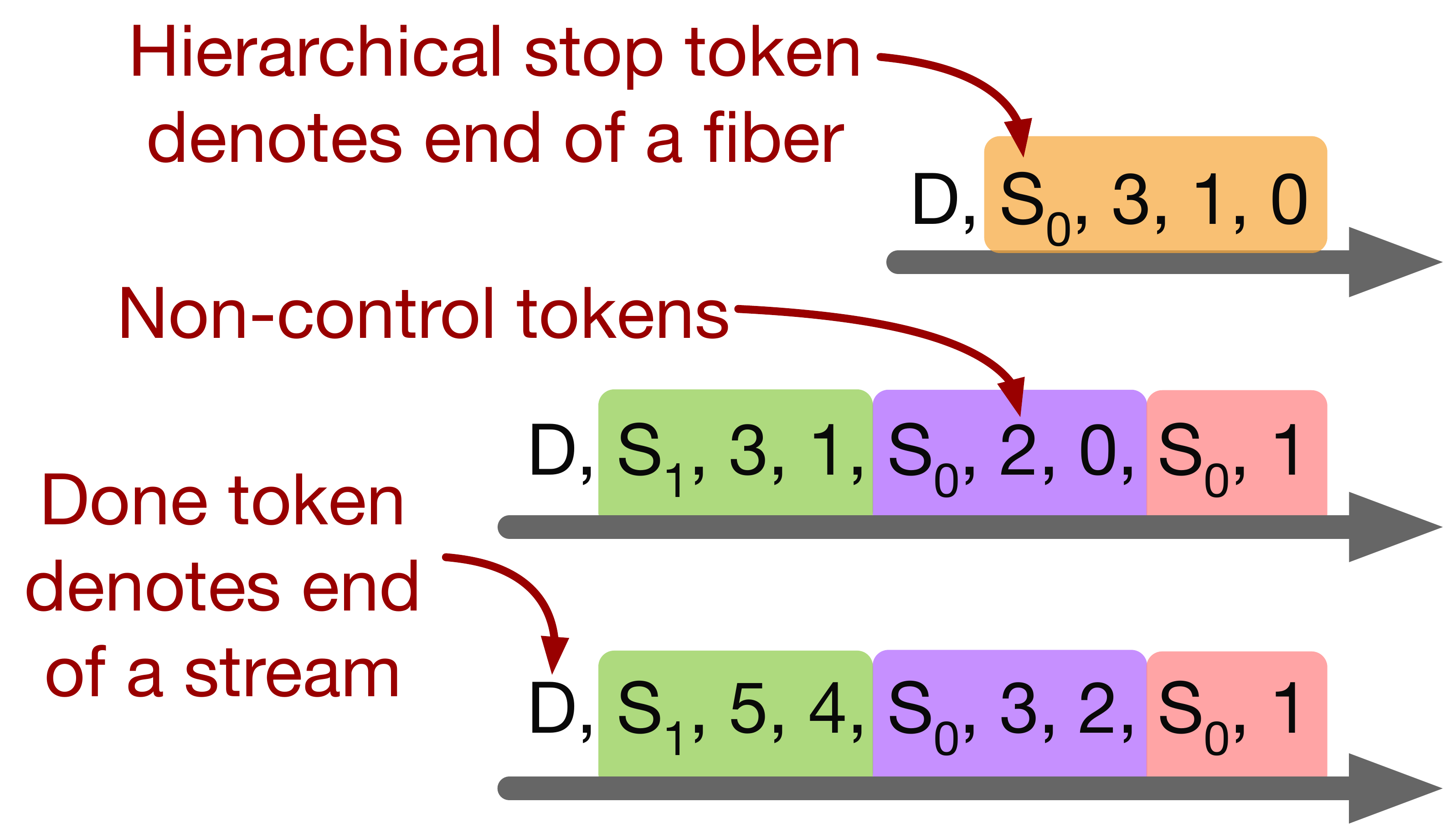}
        \label{fig:data-model-streams}
    }
    \vspace{-2mm}
    \caption{
        The data model of the SAM models sparse tensors (\Cref{fig:data-model-matrix}) as a coordinate fibertree, (\Cref{fig:data-model-coordinate-tree}) that can be stored in memory (\Cref{fig:data-model-memory}) as DCSR data structure, or sent through streams (\Cref{fig:data-model-streams}) where time increases from right to left.
        \label{fig:data-model} 
    }
\end{figure*}

\section{The Core Sparse Abstract Machine}
\label{sec:sam}

\sam{} provides a clean method to transport tensors on wires and to express all tensor algebra operations, serving as an LLVM-like interface. 
We define \samCoreBlockCount{} types of dataflow blocks that can be composed to execute arbitrary sparse tensor algebra expressions. 
\Rdscan{s} fetch a tensor's nonzero coordinates and send them as streams to intersecters, unioners, and repeaters that combine coordinates from different tensors. ALUs and reducers compute tensor operations. Coordinate droppers filter out unnecessary coordinates, and \wrscan{s} write the resulting sparse tensor to arrays in memory.

\sam{} lets programs use as many blocks as needed. Of course, any physical implementation is constrained to a finite set of resources. Our compiler can be used to transform an unconstrained graph to a specific physical backend by breaking up the computation in time through data movement into temporary memories and block reuse.

\sam{} is sufficiently expressive to represent any dataflow for any tensor algebra expression, as it implements all the features in \Cref{sec:features}. Furthermore, \sam{} implements a streaming model of those features, and Kovach and Kjolstad have shown an equivalence proof of sparse tensor algebra and a formal streaming model~\cite{kovach2022}.

\subsection{Tensor Data Model}
\label{sec:sam-data-model}

In the \sam{} abstract data model, each tensor is a coordinate tree where each tree level represents the coordinates of a different tensor dimension. This coordinate tree abstraction was first introduced as part of the TACO system~\cite{kjolstad2017tensor,chou2018} and further abstracted and formalized as \textit{\crdtree{s}}~\cite{fibertrees,sparseloop}. \Crdtree{s} are tries where each coordinate at one level is linked to a \textit{\fiber{}}---a list of child coordinates---at the next level. Crucially, only those children whose sub-trees have nonzeros are stored. \Cref{fig:data-model-matrix} shows a sparse matrix and \Cref{fig:data-model-coordinate-tree} its corresponding \crdtree{}. The matrix is stored in row-major order, so the $i$ coordinates (orange circles) are stored at the top \crdtree{} level. The $i$ coordinate 2 is not stored since its sub-tree (the third row) has only zeros. The middle level stores a $j$ coordinate for every nonzero component and the last level stores nonzero tensor values. \Crdtree{s} are useful for reasoning about tensors level by level without considering the exact storage representation. 

\Crdtree{s} are stored in memory and transmitted via streams. When in memory, each tree level is separately assigned a storage type that specifies its data representation. A level's data representation can be an uncompressed level that stores a single number encoding the \fiber{} size or it may be a compressed data structure that stores only coordinates with nonempty sub-trees. Many other data representations are possible with this abstraction~\cite{chou2018,fibertrees,sparseloop}. \figref{fig:data-model-memory} depicts one possible in-memory data structures for the \crdtree{} in \figref{fig:data-model-coordinate-tree}, where both levels are stored in compressed data structures. This storage format is called doubly-compressed sparse rows (DCSR), where a segment array denotes the start and stop reference positions of each segment in the coordinate array. A segment is one way to encode \fiber{} data associated with an array representation. Concretely in \Cref{fig:data-model-memory}, the level $j$ segment [3, 5) refers to the green level $j$ coordinates [1, 3] since the coordinates are located at indices [3, 4] in the level $j$ coordinate array.

\subsection{Tensor Streams}
\label{sec:sam-stream}

\sam{} streams are abstractions of physical wires that connect processing blocks and transmit data between these blocks.
Each \sam{} stream is a sequence of tokens that transmits one level of \crdtree{} data, along with stop tokens ($S_n$) denoting the hierarchical \fiber{} boundaries within a level, and a done token ($D$) to mark the end of a stream. There are three types of \sam{} streams: coordinate streams (abbreviated as \texttt{crd}) that transmit coordinate levels, value streams (\texttt{vals}) that transmit last-level tensor values, and reference streams (\texttt{ref}) that transmit references to the location of each coordinate's child \fiber{} in memory. \Cref{fig:data-model-streams} shows the \crdtree{} in \Cref{fig:data-model-coordinate-tree} as coordinate and value streams. Streams can be interpreted as variable-length nested lists where each stop token represents a parenthesis. Thus, the value stream in \Cref{fig:data-model-streams},
\setlength{\abovedisplayskip}{1pt}
\setlength{\belowdisplayskip}{1pt}
$$\underleftarrow{\;1, \; \text{S}_0, \; 2, \; 3, \; \text{S}_0, \; 4, \; \; 5, \; \text{S}_1, \; \text{D}\;}$$
represents the nested value level 
$$((1), (2,3),(4,5)).$$ 
The data closest to the arrowhead is sent first, and the done (D) token terminates the stream.

\begin{figure}[b]
    \centering
    \includegraphics[width=\linewidth]{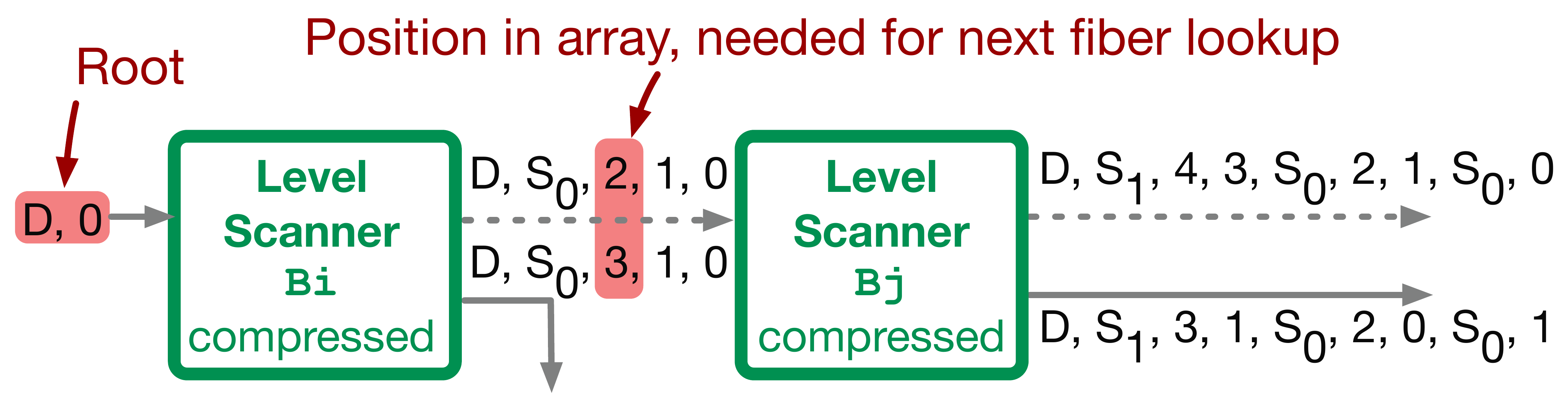}
    \vspace{-6mm}
    \caption{
        Composition of \rdscan{} blocks.
        \label{fig:block-readscanner}
    }
\end{figure}

\begin{figure}[b]
    \centering
    \includegraphics[width=\linewidth]{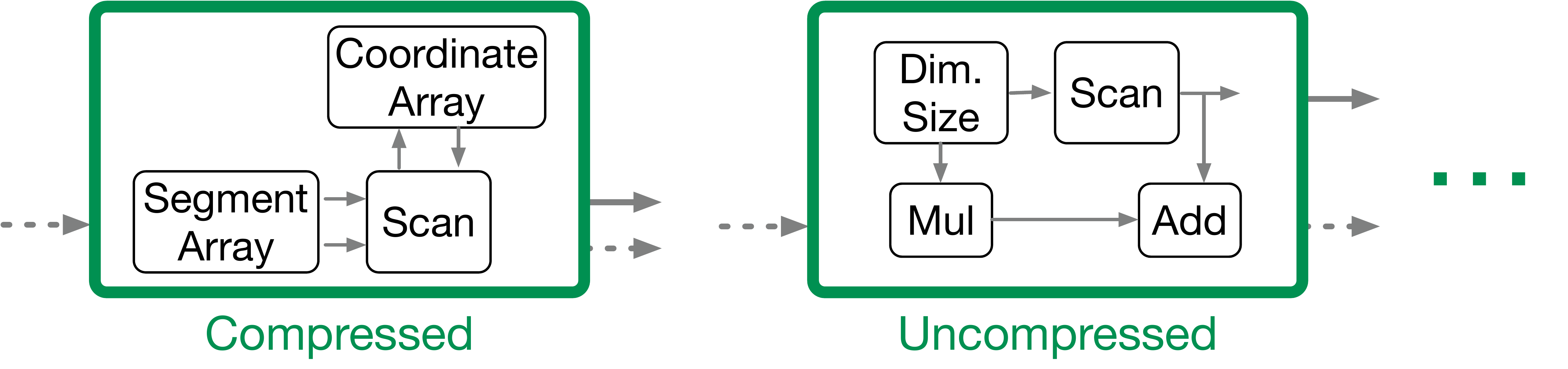}
    \vspace{-6mm}
    \caption{
        Implementations of the level scanner interface.
        %Implementations of \rdscan{s} demonstrating the format agnostic interface. %Black blocks are physical hardware. 
        \label{fig:readscanner-implementations}
    }
\end{figure}

\begin{figure*}
    \centering
    \includegraphics[width=\linewidth]{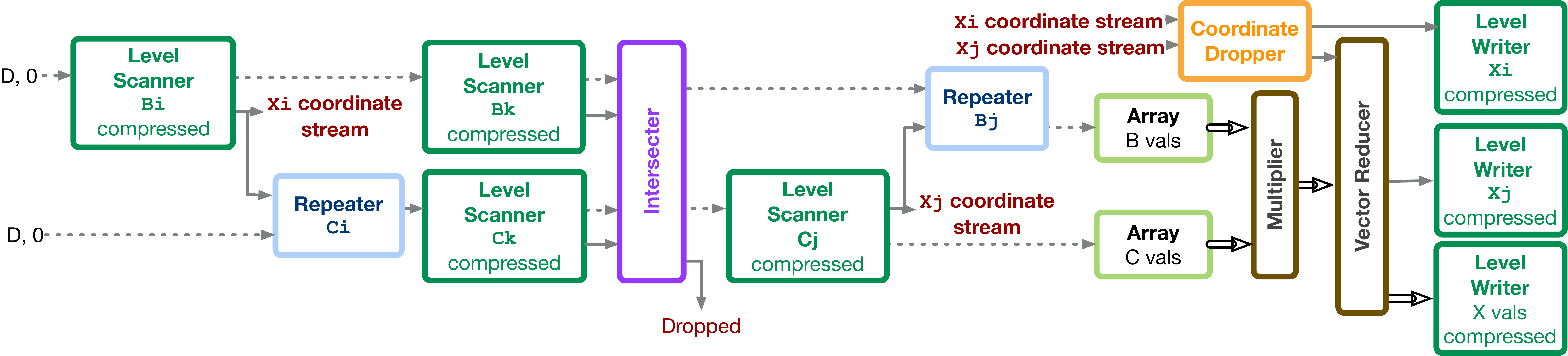}
    \vspace{-6mm}
    \caption{
        The \sam{} dataflow graph for sparse matrix multiplication $X_{ij} = \sum_k B_{ik} C_{kj},$ on DCSR matrices with linear combination of rows ($i \rightarrow k \rightarrow j$ order). Stipled, solid, and double arrows resemble reference, coordinate, and value streams respectively.
        \label{fig:sam-gemm}
        }
\end{figure*}

\subsection{Tensor Iteration}
\label{sec:sam-iteration}

\sam{} sparse dataflow algorithms start with \rdscan{s} that load tensors from memory and turn them into streams. 

\begin{block}[Level Scanner]
\label{block:rdscan}
A \rdscan{} takes in a reference stream and outputs two streams: one of coordinates and one of references. It produces a single \crdtree{} level on its output coordinate stream, \fiber{} by \fiber{}. Each \numerical{} token on the input stream is a reference to a single \fiber{} location for a given level in memory. The \rdscan{} generates all coordinates in that \fiber{}, along with their corresponding references, and then adds an additional stop token to denote the end of the \fiber{}.  
\end{block}

Each \sam{} \rdscan{} generates \fiber{s} for only one dimension. Therefore, multiple scanners must be composed to iterate over a multidimensional tensor. The composition uses the references emitted from each successive \rdscan{} to locate the \fiber{s} of the next \rdscan{}. The key to this composition is that \rdscan{s} communicate information by embedding both \fiber{} location and coordinate hierarchy---needed by downstream \rdscan{s}---into the reference streams. Each \rdscan{} adds a level to the hierarchy by either adding an $S_0$ stop token at the end of each scan or by incrementing all input stop tokens by one. They thus chain together to load an entire tensor and to convert it to per-level streams. \Cref{fig:block-readscanner} shows two \rdscan{s} that iterate over the compressed matrix in \Cref{fig:data-model-memory}. 
The reference stream emitted by the final-\rdscan{} is sent to blocks that load values from memory, as described in \Cref{sec:sam-compute}. Each \rdscan{} also connects to a memory array (\Cref{block:array}) that stores the \fiber{} and coordinate information for the level, but these are not shown in figures to reduce clutter.

The SAM \rdscan{s} support iterating over tensors stored in various in-memory level formats presented in~\cite{chou2018}, which decouples an algorithm from the tensor formats. Thus, the interfaces of the \rdscan{} are format agnostic and \figref{fig:readscanner-implementations} demonstrates how they remain unchanged as the level format implementation varies.

\subsection{Illustrative Example}
\label{sec:sam-example}

We will use the linear combination of rows algorithm (sometimes referred to as Gustavson's algorithm~\cite{gustavson1978}) for sparse-matrix sparse-matrix multiplication (\spmspm{}) to illustrate the operation of \sam{} blocks, and to demonstrate how their composition defines different algorithms. The Einstein summation notation for this algorithm is $X_{ij}=\sum_k B_{ik}*C_{kj}$, where the  matrix multiplication is accomplished by using an index order of $i\rightarrow{}k\rightarrow{}j$~\cite{zhang2021gamma}. The advantage of this iteration order is that $k$ coordinates are first intersected and only those $k$s that survive result in further computation.

\Cref{fig:sam-gemm} shows the algorithm as a \sam{} dataflow graph. From the left, the coordinates of the two matrices are loaded from DCSR data structures in memory by \rdscan{s}. The coordinates are then transformed into a three-dimensional iteration space by chaining together the $i\rightarrow{}k$ coordinates of the $B$ matrix with the $k\rightarrow{}j$ coordinates of the $C$ matrix. This space requires duplicating data to fill in missing dimensions.  In this example each matrix is broadcast over an index variable of the other matrix ($B$ over $j$ and $C$ over $i$).

\subsection{Stream Merging} 
\label{sec:stream-merge}

Once the operand coordinate streams have been generated, the next task is to merge them. The index variables of a tensor index notation expression create an iteration space that we must cover, taking advantage of both the sparsity of the tensors and the mathematical properties of the operations to avoid unnecessary computation. Our design covers this sparse iteration space hierarchically by merging the coordinates of one dimension at a time, with the surviving coordinates from one dimension dictating what \crdtree{} \fiber{s} need to be merged in the next dimension. The hierarchical merging is implemented with per-level merging blocks (intersection and union) and repetition machinery to handle the case where a tensor is broadcast~\cite{iverson1962programming,harris2020array} across the dimension of another tensor, as required by our illustrative example in \Cref{fig:sam-gemm}.

The merging operations combine $m$ streams, representing the same coordinate level of all operand tensors, \fiber{} by \fiber{}. Coordinate merging is inherently a set operation: specifically, intersection (since $a \cdot 0=0$) and union (since $a+0=a$) suffice for tensor algebra.

\begin{block}[Intersecter]
\label{block:intersect}
    An intersecter has $m$ pairs of coordinate and reference streams go in and one coordinate stream and $m$ reference streams come out. It outputs coordinates and corresponding input references when all input coordinates are equivalent. 
\end{block}

\begin{block}[Unioner]
\label{block:union}
   A unioner has the same input/output interface as the intersecter. However, it outputs coordinates and their associated input references whenever there exists at least one coordinate from any input. If the coordinate exists only on $p$ inputs where $p < m$, the union block outputs an \ntkn{} ($N$) token on the the other $m-p$ output reference streams.
\end{block}

\begin{figure}[b]
    \centering
    \includegraphics[width=\linewidth]{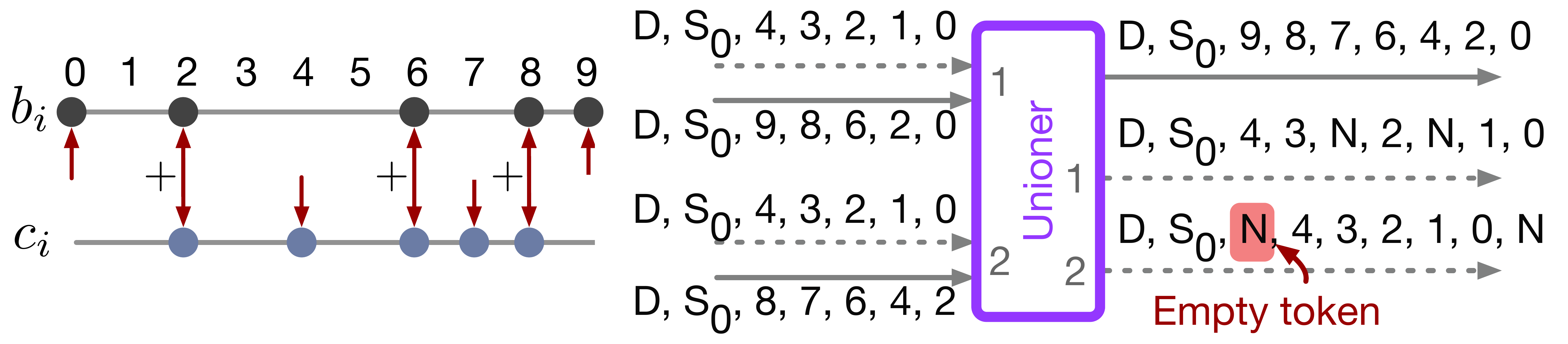}
    \vspace{-6mm}
    \caption{
        Example of union coiteration for $b_i+c_i$
        \label{fig:union}
    }
\end{figure}

\Cref{fig:union} shows an example of a binary unioner that produces a coordinate stream that is the union of two input streams, along with the references from each input reference stream whose coordinates survived the union. Both emitted reference streams are augmented with \ntkn{} tokens (N) to have the same shape as the emitted coordinate stream.

\begin{figure}
    \centering
    \includegraphics[width=\linewidth]{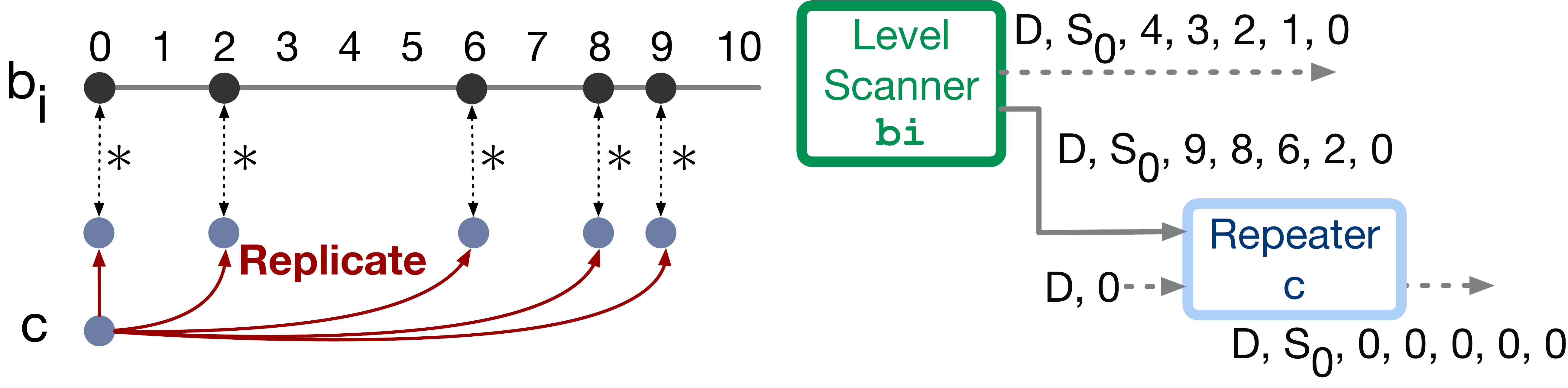}
    \caption{
        Repeating a scalar with a \repeater{} block.
        \label{fig:repeat-mechanism}
    }
\end{figure}

As we saw in \figref{fig:sam-gemm}, it is common for expressions to replicate one tensor across a dimension of another, often called array broadcasting. \Cref{fig:repeat-mechanism} shows a simple vector scaling example. It demonstrates how the repeater block replicates a reference stream over every coordinate of the provided coordinate stream. The repeater is a new primitive that solves limitations on prior work architectures needing to pre-configure higher-order iteration counts~\cite{spu, hegde2019extensor, rucker2021capstan}.

\begin{block}[\xmakefirstuc{\repeater{}}]
\label{block:repeat}
Repeaters have one input coordinate stream, one input reference stream, and one output reference stream. Each \numerical{} token in the input reference stream is repeated $m$ number of times, where $m$ is the number of \numerical{} tokens from the input coordinate stream before a stop token is seen. 
\end{block}

Hierarchical repeating and stream merging compose to express algorithms for multidimensional tensor contractions. Reconsider the linear combination of rows \spmspm{} algorithm from \Cref{fig:sam-gemm}. The $i$ coordinates loaded from $B$ are not only passed to the $i$ level writer of $X$ by way of the coordinate dropper, but also fed to a repeater that broadcasts all of $C$'s $k$ coordinates over each $i$. 

\subsection{Computation}
\label{sec:sam-compute}

After stream merging, the remaining coordinates are coordinate space points that contribute to the result. 
Their corresponding reference streams are passed to array blocks that load their values.

\begin{block}[Array]
\label{block:array}
An array block is a proxy for a memory interface and can be treated  as a contiguous section of memory. It has two interface modes---load, which given one input reference stream fetches data to produce one output  stream of any type, and store, which given one input reference stream and one input data stream of any type has a side effect that stores the data to its corresponding reference location in memory.
\end{block}

In \sam{}, arrays store values, coordinates, and references. In the computation pipeline, value streams are read from the array of each operand, with the same coordinates, and combined using streaming arithmetic-logic units (ALUs). The \Cref{fig:sam-gemm} ALU is a multiply unit.

\begin{block}[ALU]
\label{block:add}
An ALU block consumes two value streams and produces one value stream. It applies an arithmetic operator (add, subtract, or multiply) to inputs, treating \ntkn{} tokens as zeros.
\end{block}

In addition to combining values at the same coordinate, often the algorithm needs to accumulate a tensor. In our illustrative example in \Cref{fig:sam-gemm} this occurs at the end, where we sum over the $k$ dimension (multiple dimensions can be summed by chaining reduction blocks). Reductions in tensor algebra may occur over any tensor dimension, independent of the order in which we choose to merge coordinates.  Thus, summation reductions may occur over the coordinate level that is merged last (requiring a scalar to accumulate the result), over the coordinate-level merged second to last (requiring a vector to accumulate the results), or over coordinates merged earlier (requiring a higher-dimensional tensor to accumulate the results). \sam{} provides one block for reductions that must be configured for any specific dimension of accumulation.

\begin{block}[Reducer]
\label{block:reduce}
A reducer is configured by $n$, the dimension of the memory needed in the reduction. It inputs and outputs $n$ coordinate streams and one value stream. The block is sent an entire $n$-dimensional (sub-)tensor with repeated points/values and outputs streams that represent that tensor with unique coordinates and summed values. Specific reducers include: scalar where $n=0$, vector where $n=1$, and matrix where $n=2$.  
\end{block}

The reducer internally adds values corresponding to equivalent coordinate points and stores the results in an internal storage, which may be a dense or a sparse data structure. Finally, when an $n$-level reduction is completed, for example when a whole row has been processed for the Gustavson's algorithm in \Cref{fig:sam-gemm}, the reducer emits the resulting tensor as streams with deduplicated coordinates. When accumulating coordinates with empty fibers, resulting from ineffectual intersections, the reducer may be configured to either accumulate empty fibers into an explicit zero (the identity for addition) or to remove the empty fibers by removing their extra stop tokens. The choice is an implementation decision, but empty reduction behavior may affect \sam{} graph construction for other blocks (see \Cref{block:crd-drop} and \Cref{tab:primitive-study}).

\begin{figure}
    \centering
    \includegraphics[width=0.82\linewidth]{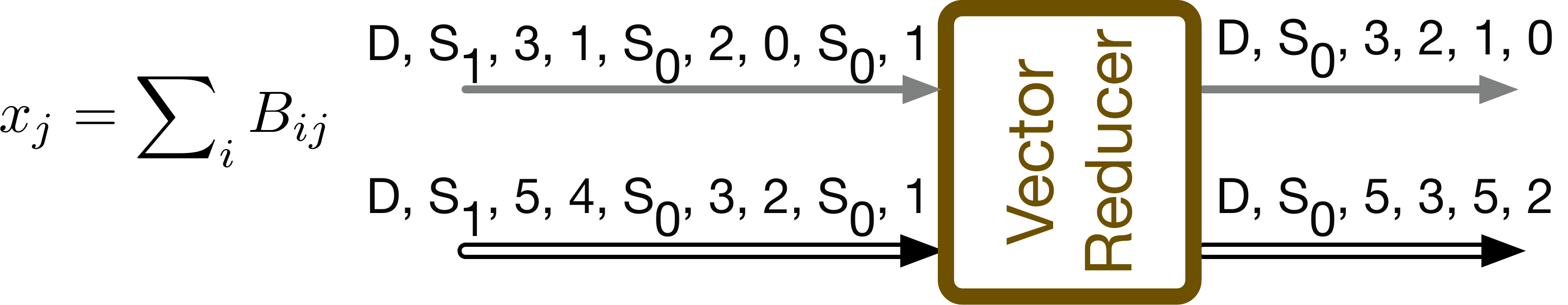}
    \caption{
        Example using the row reducer, where $n=1$, to accumulate the columns of the matrix from \Cref{fig:data-model-matrix}.
        \label{fig:reducer-mechanism}
    }
\end{figure}

Like with \rdscan{s}, various implementations of the reducer are possible underneath the abstraction, including k-way merging, dense arrays, compressed data structures, and bitmaps~\cite{rucker2021capstan, phi2019mukkara, plasticine}. \figref{fig:reducer-mechanism} shows an example of a row ($n=1$) reducer.

\subsection{Tensor Construction}
\label{sec:sam-lhs}

The final step of a \sam{} graph is to store the resulting tensor streams back to memory. Specifically, the surviving coordinate streams for the index variables used to index the left-hand side of the Einsum expression, as well as the computed values, need to be stored back into per-level tensor memory representations. 

\begin{block}[\xmakefirstuc{\wrscan{}}]
\label{block:wrscan}
\xmakefirstuc{\wrscan{s}} take in either one value stream or one coordinate stream and store its contents to memory, internally generating reference information and auxiliary level data structures. As a result, the block is a wrapper around the store mode of a coordinate Array (and its metadata) or a value array. The \wrscan{'s} internally generated references store the data tokens from the input stream in order. 
\end{block}

\begin{figure}
    \centering
    \includegraphics[width=\linewidth]{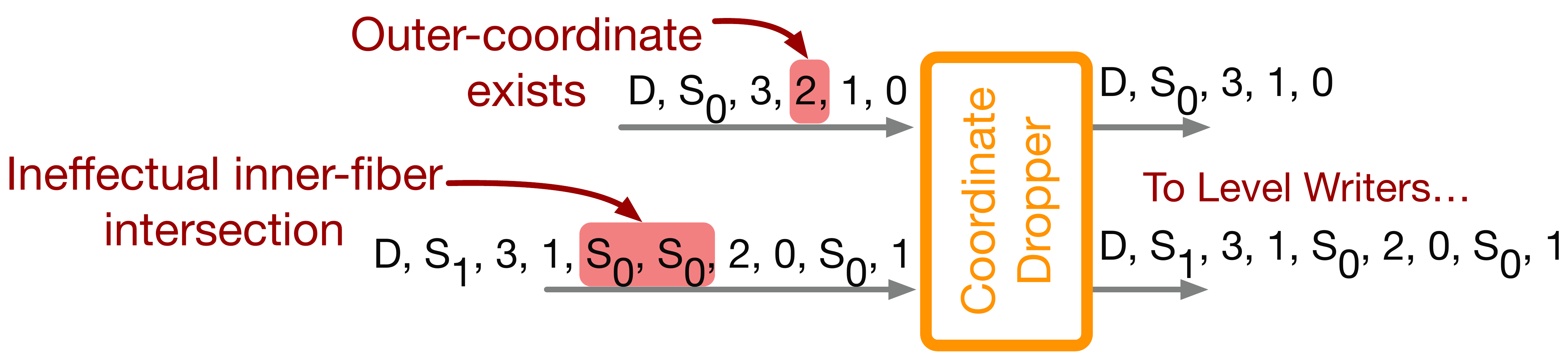}
    \vspace{-5mm}
    \caption{
        Dropping coordinate 2 from the matrix in \Cref{fig:data-model-matrix}.
        \label{fig:crd-drop-mechanism}
        \vspace{-6mm}
    }
\end{figure}

In cases with at least one index-variable level above an intersection level, the result coordinate streams must be cleaned before the \wrscan{} stores it back to memory. The coordinate cleanup removes any outer-level result coordinates that have ineffectual inner-level intersections (either empty intersections or zero values) as shown in \figref{fig:crd-drop-mechanism}. We introduce the coordinate dropper block to handle these cases. Coordinate droppers with value stream inputs are optional if explicit zeros need not be removed. In this case, other coordinate dropper blocks are also optional if the reducer blocks is configured to accumulate empty fibers into 0-values, since ineffectual (empty) intersections will produce explicit values. 

\begin{block}[Coordinate Dropper]
\label{block:crd-drop}
The coordinate dropper takes in one outer-level coordinate stream and one inner-level coordinate or value stream.
It removes both the outer-level and inner-level tokens that came from ineffectual merging or computation (empty \fiber{s} or zeros) at the inner level.
\end{block}

\subsection{Alternatives and Tradeoffs Discussion}
We discuss alternatives and tradeoffs of core SAM design decisions, which we hope will provide insights into our design rationale.

\subsubsection*{Stream Control Tokens}
\label{sec:control-tradeoff}
\Cref{sec:sam-data-model} presents a solution for streaming dataflow where the control tokens (stop, empty, and done) are directly passed through the data plane, but alternative solutions to control flow for dataflow exist. Other valid dataflow control-flow solutions include control signaling on dedicated control-only streams, embedding control (via counters and control tokens) directly into each primitive, and having a completely separate control plane. We give examples, discuss the limitations of these approaches, and justify putting control tokens on the data plane.

In initial iterations of SAM, we considered representations with primitives that had dedicated control signaling using control-only streams. For example, we considered passing repeat information by connecting two level scanners together to exchange signals that denote when to stop repeating coordinates. Having streams that only communicate control information risks underutilization, where no information is passed through most cycles, and complicates the composition of multiple blocks, as they would need to be hardened together. However, the benefit of this design would be that control information (when produced) and data can be processed in parallel. 

We also considered a SAM design that embedded control directly into each block, which included embedding repeat counters into level scanners and done signaling into each primitive. The direct embedding of counters and other control logic into the primitives increases the primitive area and hardware complexity. Additionally, pre-configuration of counters is usually necessary, meaning that metadata information (like number of nonempty elements) must be obtained by the compiler statically during compile time by iterating over the sparse data at least once on the CPU.

Finally, a design with a separate control plane (not just separate control wires) would be more similar to a von Neumann architecture than prior work on dataflow architectures. Dataflow architectures attempt to remove performance overheads of traditional CPUs by eliminating most general-purpose control, which is done by removing the control unit and restricting control. Having a separate control plane on SAM would fundamentally push our design closer to a von Neumann machine abstraction and would thus not be a good fit for representing streaming dataflow accelerator backends.

Although having control tokens directly processed on the data plane may decrease performance---primitives must now process these tokens---they increase interconnect and logic utilization, decrease primitive area, minimize primitive logic complexity, and still allow for streaming dataflow processing (massive pipelining/parallelism with low control overhead).

\subsubsection*{Level-Based Stream Representation}
\label{sec:point-streams}
Another approach to our level-based streaming tensor representation would be a less efficient point-based streaming representation. One implementation could stream flattened tensor point tuples with no control tokens. The tensor from \Cref{fig:data-model} could thus be represented as  
$$\underleftarrow{(0, 1, 1),\; (1, 0, 2),\; (1, 2, 3),\; (3, 1, 4),\; (3, 3, 5),\; D}$$
In this representation, the number of processed stream tokens for identity matrices is $3 \cdot \text{nnz}_B$, where $\text{nnz}_B$ is the number of nonzeros in $B$. 
We can compare the two representations to find when the point-based representation has more tokens using the equation $3 \cdot \text{nnz}_B > (1 + c) \cdot \text{nnr}_B + 2 \cdot (1+c) \cdot \text{nnz}_B$ where $c$ is the fraction of control tokens and $\text{nnr}_B$ is the number of nonempty rows in $B$. Using worst-case numbers from our analysis in \Cref{fig:eval-stream-overhead}, we rewrite the equation to $3 \cdot \text{nnz}_B > 1.3326 \cdot \text{dim}_{B_i} + 2 \cdot 1.3326 \cdot \text{nnz}_B \implies \text{nnz}_B > 3.98 \cdot \text{dim}_{B_i}$ where $\text{dim}_{B_i}$ is the number of rows in $B$. The result demonstrates that our level-based representation, in the worst-case, processes less tokens than the point-based approach when there are on average more than 4 elements per row. Of the matrices we selected in~\Cref{fig:eval-stream-overhead}, all 5 middle 50 and 5 large 50 matrices satisfy the $4\times$ inequality and are more efficient in our level-based representation.
Our approach becomes even more efficient for higher-order tensors. The coordinates at every level are expanded to the last level---proportional to roughly O($n^{N}$) instead of O($n^2$) for matrices, where $n$ is a single tensor dimension and $N$ is the tensor order---to produce the tensor point tuples.

\section{Optimization Discussion}
\label{sec:optimizations}

The core \sam{} blocks introduced in \Cref{sec:sam} are complete in the sense that they compose to express every tensor algebra expression. Moreover, they suffice to express all coordinate processing (dataflow) orders and fusion---the primary tools to construct algorithms with good asymptotic complexity~\cite{ahrens2022}.
To express \sam{} graphs that further optimize performance and deal with finite hardware, we have added additional capabilities. These capabilities let the graphs express parallelism, tiling, and more ways to represent tensor information either in memory or as streams. In this section, we discuss how \sam{} extends to include these additional optimizations and how they compose with the core \sam{} from \Cref{sec:sam}.

\subsection{Tiling}
\label{sec:tiling}

\begin{figure}
    \centering
    \includegraphics[width=0.9\linewidth]{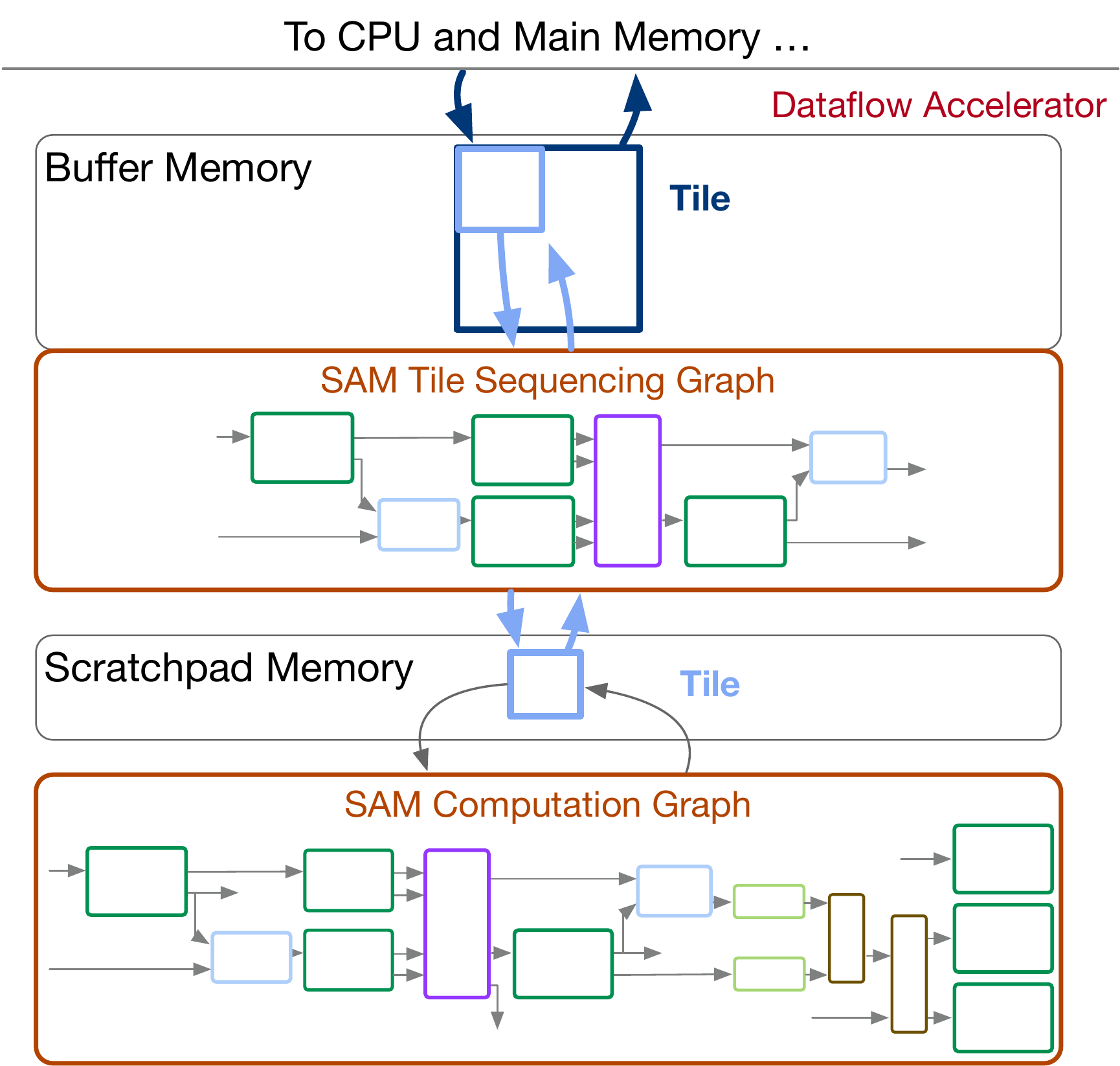}
    \caption{How to sequence tiled tensors (shown in blue) to fit in finite memory for \spmspm{}. The SAM computation graph is the same as in \Cref{fig:sam-gemm}, and the SAM tile sequencing graph performs coiteration and merging of tile coordinates.}
    \label{fig:tiling}
\end{figure}

In our data model, tiling a tensor splits a single \crdtree{} level into multiple levels and then reorders those levels to produce smaller sub-tensors (tiles). \Cref{fig:tiling} shows how \sam{} can sequence tiled tensors between host and accelerator devices for computation with fixed-size memories. \sam{} graphs are used in outer levels to sequence the tile coordinates (tile IDs) for reuse and in the inner levels to perform the computation. The tile sequencing is equivalent to tensor iteration (\Cref{sec:sam-iteration}) and stream merging (\Cref{sec:stream-merge}), where tile IDs are coordinates and the values are references to the next level of tiles. As in ExTensor~\cite{hegde2019extensor} and Capstan~\cite{rucker2021capstan}, we assume that tensors are tiled beforehand so that each tile fits in the dataflow accelerator's memory hierarchy. We demonstrate in \Cref{sec:eval-model} \sam{}'s ability to fit tensors into finite memories and the tradeoff space of different memory configurations (dictated by architectural and implementation-specific memory configurations, like the maximal tile size and bandwidth at each level of the memory hierarchy). 

\subsection{Tensor Locating}
\label{sec:locate}

In \Cref{sec:sam}, all intersections are performed using coiteration, where the coordinates are intersected using a two-finger merge strategy. This is sufficient for computational correctness, however, it can be asymptotically inefficient. (The core \sam{} has to coiterate between an uncompressed dense counter level and a compressed level even though it is sufficient to iterate through just the sparse level.) We can often improve intersection efficiency if one tensor has far fewer elements than the other. Rather than waiting for the larger tensor to stream all its level coordinates, it can be more efficient to ask the larger tensor if it contains any of the coordinates from the smaller tensor. This operation, known as iterate-locate or leader-follower intersection, is possible with another \sam{} block that uses a coordinate instead of a reference to index an array.

\begin{block}[Locator]
\label{block:locate}
A locator takes in one coordinate and reference stream and outputs one coordinate and two reference streams. For each coordinate, the block finds the associated reference within an array block, if it exists, and outputs that reference and the input coordinate and reference. Otherwise, it emits an empty \fiber{} on all streams.
\end{block}

With locators, we can reorganize \sam{} graphs to remove intersecters. A prominent example that benefits from this optimization is the inner product sparse matrix-vector multiplication, where the vector is dense. By streaming through the coordinates of each matrix row and locating into the vector, we avoid loading the values of the vector whose corresponding matrix value is zero. Locate blocks can also be used to scatter into a result that supports random insert, such as a dense left-hand-side tensor. Thus, the linear combination of rows matrix-vector multiplication can avoid a vector reducer.

\label{sec:crd-skip}

Locators can also speed up intersection when used in conjunction with intersecters that communicate information back to \rdscan{s} about coordinate ranges that are no longer needed. This optimization, called coordinate skipping or galloping, is common in software and has also been proposed in hardware~\cite{hegde2019extensor}. 
In coordinate skipping, the intersecter sends a signal back to the trailing \rdscan{} (extending the interface of both blocks from the definitions in~\Cref{sec:sam}), informing it of the coordinate that is needed next. The \rdscan{}, in conjunction with a locator, then skips ahead to this coordinate and avoids sending useless coordinates between its current coordinate and the coordinate sent by the intersecter.

\subsection{Bitvectors}
\label{sec:bitvectors}
Bitvectors are a natural way to compress coordinate information since bits are easy to implement in hardware. Bitvectors have a $1$ in positions where explicit coordinates exist and a $0$ for empty (or zero) coordinates. Bitvectors have a pseudo-dense iteration space---one that iterates proportional to some constant factor of the dense dimension of each tensor level. This iteration is usually asymptotically worse in performance when compared to compressed iteration, especially with increasing sparsity. However, bitvectors may also increase efficiency since an $n$-bit bitvector, encoding $n$ coordinate elements, can be processed in one cycle.
In some prior-work hardware like Capstan~\cite{rucker2021capstan} and SIGMA~\cite{qin2021sigma}, bitvectors are the only compression protocol. Bitvectors may also be offered in addition to compressed coordinates, with blocks that convert between their stream protocols. We introduce a bitvector converter that transforms a coordinate stream into a new bitvector stream protocol and describe a \rdscan{} for the bitvector level format.  

\begin{block}[Bitvector Converter]
\label{block:bitvector}
Bitvector converters transform $b$ coordinates from the input coordinate stream into a single bitvector token of $b$ bits on the bitvector stream output. Each bit indicates whether it has children or whether its sub-tree is empty.
\end{block}

The \sam{} bitvector \rdscan{} is similar to \Cref{block:rdscan}, but it outputs a bitvector stream instead of a coordinate stream. The bitvector \rdscan{} also changes the reference stream behavior presented in \Cref{sec:sam-stream}. Consider the $b$ vector from \Cref{fig:repeat-mechanism} that produces the coordinate stream $\underrightarrow{D, S_0, 9, 8, 6, 2, 0}$ compressed as the 4-bit bitvector stream  $\underrightarrow{D, S_0, 0011, 0100, 0101}$. As a reminder, the compressed \rdscan{} would output a reference stream of $\underrightarrow{D, S_0, 4, 3, 2, 1, 0}$ to indicate contiguous references (positions) in memory. However, the bitvector \rdscan{} instead produces the reference stream $\underrightarrow{D, S_0, 3, 2, 0}$ that sums bitcounts (popcounts) to find the positions in memory for the next level. The bitvector format thus demonstrates how \sam{} handles various stream types as different compression protocols on the wires (coordinates versus bitvectors), along with various reference stream protocols, while maintaining composability.

\subsection{Parallelization}
\label{sec:parallel}

Given the spatial streaming abstraction in \sam{}, parallelism is easily representable via vectorization and graph duplication. Conceptually the simplest extension is to vectorize streams as wire buses and to update the blocks to handle the increased data rates.  

To enable coarse-grained parallelism, \sam{} dataflow graphs can fork streams with a parallelizer and join streams with a serializer. The parallelizer block takes in a sequential tensor stream and parcels out different elements to multiple output streams concurrently. The serializer block works inversely and joins parallel streams into a sequential stream by interleaving their coordinates.

\section{The Custard Compiler}
\label{sec:compiler}

\begin{figure}[htb]
    \centering
    \includegraphics[width=\linewidth]{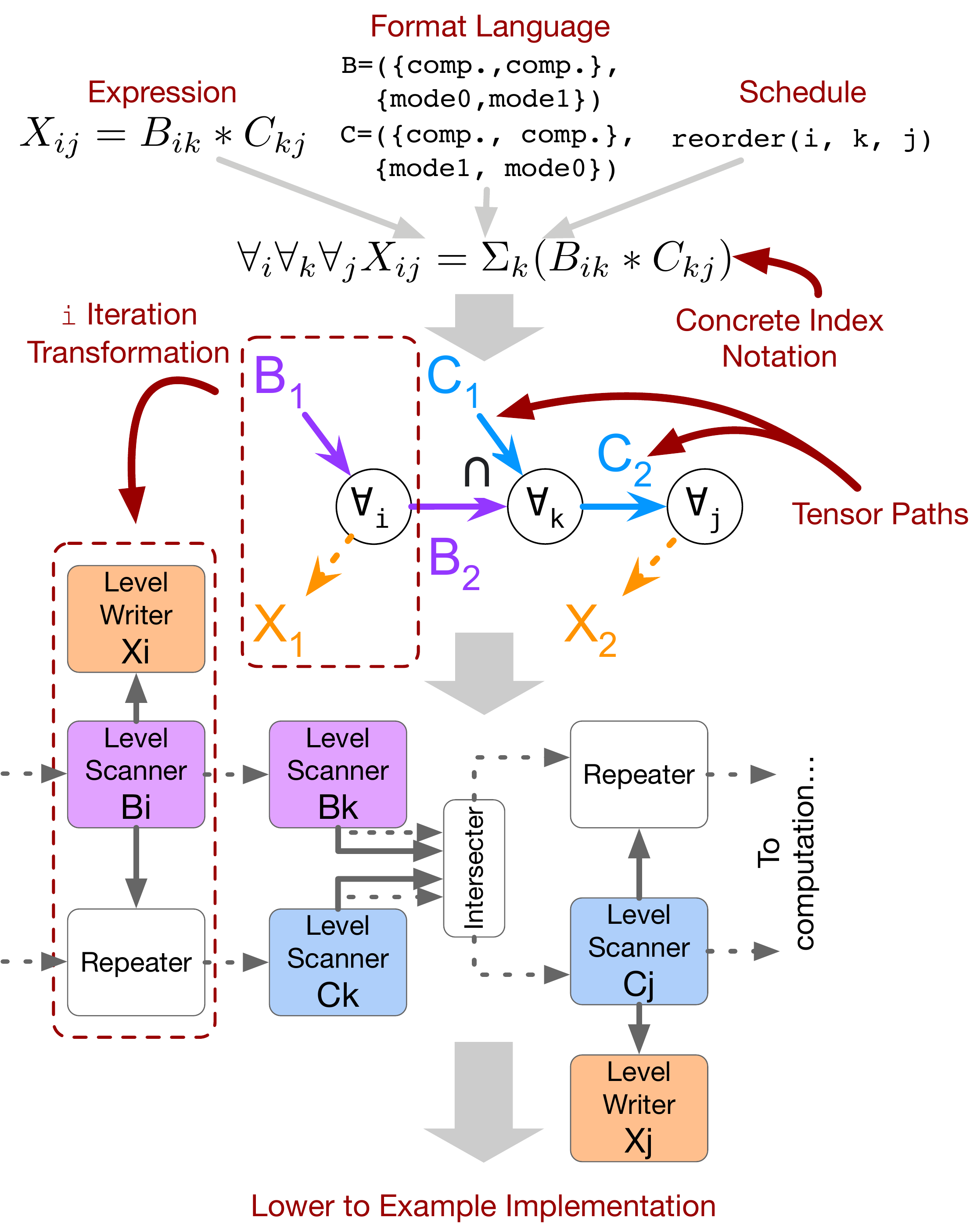}
    \caption{Custard's steps for compiling \sam{} tensor iteration, merging, and construction for the \spmspm{} example in~\Cref{sec:sam-example}. We abbreviate compressed as comp. From top to bottom, Custard uses the TACO input APIs to generate concrete index notation, creates index-variable paths for each tensor, and constructs the partial \sam{} graph (where the color of each block corresponds to a tensor path---purple for $B$, blue for $C$, and orange for $X$). Custard lowers the dotted red region of the tensor paths to the dotted red region of \sam{} blocks. 
    \label{fig:compiler}}
\end{figure}

The Compiler for Unified Sparse Tensor Algebra Reconfigurable Dataflows (Custard) is our compiler to \sam{} which acts as an intermediate representation. Custard compiles tensor algebra expressions with associated data structure specifications~\cite{chou2018} and schedules~\cite{kjolstad2019,senanayake2020} to \sam{} dataflow graphs (see \Cref{fig:compiler}). Custard is an open-source C++ project that utilizes the TACO front-end~\cite{kjolstad2017tensor} but supplies a new lowerer and code generator. Custard uses TACO's three input APIs (tensor index notation, a format language, and a scheduling language) and the code that transforms these into a high-level IR called concrete index notation---an abstract loop nest with scheduling information shown in \Cref{fig:compiler}. Although Custard generates \sam{} dataflow graphs, automatic binding to prior-work hardware backends described in \Cref{sec:classification} is left as future work.

\Cref{fig:compiler} illustrates a partial compilation to the \sam{} dataflow graph for the $ikj$-order \spmspm{} example from \Cref{sec:sam-example}. Custard converts the concrete index notation to a graph that represents each tensor's path through the index variables (shown as colored arrows with tensor labels in \Cref{fig:compiler}). Custard then builds the following three sections in order: tensor iteration and merging, computation, and tensor construction. It builds the tensor iteration and merging by iterating over the Cartesian product of index variables and input tensors, which in our example is $\{i , k, j\}$ by $\{B,C\}$. For every index variable in a tensor's path, Custard places and connects a \rdscan{}, which we color corresponding to its associated tensor path. For every index variable absent from a tensor's path that does not have an outer index variable reduction, Custard inserts a repeat block. Finally, if multiple tensor paths exist for an index variable, then Custard inserts an intersecter (for multiplication) or unioner (for addition).
Next, the output reference streams from the first part are connected to the compute tree, which consists of scalar operations and reductions, (extracted from the concrete index notation). Finally, the output values from the computation section and each index variable's final coordinate stream are connected to the output construction blocks (denoted by the orange in \Cref{fig:compiler}) with coordinate drop blocks inserted as necessary.

\section{Evaluation}
\label{sec:eval}

We use Custard to compile disparate sparse tensor algebra algorithms into \sam{} graphs that are automatically lowered to a cycle-approximate functional simulator. 
The Custard code, used to automatically compile \sam{} graphs, is compiled using GCC 9.4.0. The \sam{} lowering and \sam{} simulator are written in Python 3.8. Our SAM simulator tracks each cycle iteration and models SAM graphs as fully pipelined (i.e. every primitive produces one token each cycle). It is cycle approximate since we assume for this section---except \textit{Modeling Hardware with Finite Constraints} in~\Cref{sec:eval-model}--- that: input queues are infinite, data fetched from arrays (memory) take only one cycle, memories are pre-initialized, and primitives are not time-shared. These assumptions do not affect our evaluation conclusions since this section only contains comparisons against simulator cycles.
All benchmarks are run on a 2.2 GHz Intel Xeon Silver 4214 24-core CPU with a 16 MB LLC running Ubuntu 18.04.

\subsection{Empirical Study of the Generality of SAM }
\label{sec:eval-generality}

\newcommand{\greencheck}{\color{green}\checkmark}
\newcommand{\redx}{\color{red}\scalebox{1.2}{\textbf{$\times$}}}

\begin{table*}[t]
    \caption{
    SAM primitive counts for a wide range of real-world expressions, demonstrating the expressiveness and generality of SAM. Each expression also contains a breakdown of the sparse tensor algebra features it contains. All features and primitive counts are obtained assuming the expression uses an alphabetical dataflow index-variable ordering except in \spmspm{}.\textsuperscript{b}
    \label{tab:generality-study-taco}
  }
  \begin{minipage}{\linewidth}
  \centering
  \begin{minipage}{0.9\linewidth}
  \footnotesize
  \setlength{\tabcolsep}{2.8pt}
  \begin{tabu}{ll|cccccc|ccccccccc}
  \toprule
    \rowfont{\sffamily\bfseries} 
    \multirow{3}{*}{\sffamily\bfseries{Name}} & 
    \multirow{3}{*}{\sffamily\bfseries Expression} & \multicolumn{6}{c}{Sparse Tensor Algebra Feature} & \multicolumn{9}{|c}{SAM Primitive Composition (count)} \\
    \cmidrule{3-17} \cmidrule{10-17}
    \rowfont{\sffamily\bfseries}
    & &  Out   & Input    & Num.  & Reduce   & Broad- &   
    \multirow{2}{*}{\sffamily\bfseries Op} & Lvl & Rep- & Inter- & Uni- & \multirow{2}{*}{ALU} & Red- & Crd
    & Lvl & Arr-  
   \\
    \rowfont{\sffamily\bfseries}   
    &  &  order & order & inputs & order & cast &  
    & Scan & eat & sect & on &  & uce & Drop~\textsuperscript{a} & Wr & ay\\
    \midrule
    \sffamily\bfseries SpMV         & $x_{i} = \sum_j B_{ij}c_j$         
    & 1 & 1,2 & 2 & 0 & \greencheck & *     &
    3 & 1 & 1 &   & 1 & 1 & 1 & 2 & 2 \\    
    \sffamily\bfseries \spmspm{}      & $X_{ij} = \sum_k B_{ik}C_{kj}$    
    & 2 & 2   & 2 & 0--2~\textsuperscript{b} & \greencheck & *     & 
    4 & 2 & 1 &   & 1 & 1 & 0--2~\textsuperscript{b} & 3 & 2 \\
    \sffamily\bfseries SDDMM    & $X_{ij} = \sum_k B_{ij}C_{ik}D_{jk}$   
    & 2 & 2 & 3 & 0 & \greencheck & *   & 
    6 & 3 & 3 &   & 2 & 1 & 2 & 3 & 3 \\
    \sffamily\bfseries InnerProd & $\chi = \sum_{ijk}B_{ijk}C_{ijk}$     
    & 0 & 3 & 2 & 0 & \redx & * & 
    6 &  & 3  &   & 1 & 3 &  & 1 & 2 \\
    \sffamily\bfseries TTV        & $X_{ij} = \sum_kB_{ijk}c_k$          
    & 2 & 1,3 & 2 & 0   & \greencheck & *   & 
    4 & 2 & 1 &   & 1 & 1 & 2 & 3 & 2 \\     
    \sffamily\bfseries TTM        & $X_{ijk} = \sum_lB_{ijl}C_{kl}$      
    & 3 & 2,3 & 2 & 0 &  \greencheck & *    & 
    5 & 3 & 1 &   & 1 & 1 & 3 & 4 & 2 \\
    \sffamily\bfseries MTTKRP & $X_{ij} = \sum_{kl}B_{ikl}C_{jk}D_{jl}$  
    & 2 & 2,3 & 3 & 0 & \greencheck & * & 
    7 & 5 & 3 &   & 2 & 2 & 3 & 3 & 3 \\
    \sffamily\bfseries Residual     & $x_{i} = b_i - \sum_j C_{ij}d_j$   
    & 1 & 1,2 & 3 & 0 & \greencheck & *,-   & 
    4 & 1 & 1 & 1 & 2 & 1 & 1 & 2 & 3 \\
    \sffamily\bfseries MatTransMul & $x_{i}=\sum_j\alpha B_{ij}^Tc_j+\beta d_i$ 
    & 1 & 0-2 & 5 & 1 & \greencheck & *,+ & 
    4 & 4 & 1 & 1 & 4 & 1 & 1 & 2 & 5 \\
    \sffamily\bfseries MMAdd   & $X_{ij} = B_{ij} + C_{ij}$              
    & 2 & 2   & 2 & \redx & \redx       & + & 
    4 &   &   & 2 & 1 &   &   & 3 & 2 \\ 
    \sffamily\bfseries Plus3     & $X_{ij} = B_{ij} + C_{ij}+ D_{ij}$    
    & 2 & 2 & 3 & \redx & \redx & +      & 
    6 &   &   & 2 & 2 &   &   & 3 & 3 \\ 
    \sffamily\bfseries Plus2     & $X_{ijk} = B_{ijk}+C_{ijk}$           
    & 3 & 3 & 2 & \redx & \redx & + &  
    6 &   &   & 3 & 1 &   &   & 4 & 2 \\
    \bottomrule
  \end{tabu}%\vfill
  \end{minipage}\hfill
  \begin{minipage}{0.9\linewidth}
  \footnotesize
  \textsuperscript{a} Coordinate dropper primitive counts in the \sam{} graphs assume that the reducer is configured to filter out and remove reductions of empty fibers. \\
  \textsuperscript{b} In \spmspm{} we show the features and primitive counts for all dataflow orderings: inner product, linear combination of rows, and outer product.
    \end{minipage}
  \end{minipage} %\hfill
\end{table*}
    
We demonstrate the generality of \sam{} by generating dataflow graphs for a wide range of useful sparse tensor algebra expressions, shown in \Cref{tab:generality-study-taco}, including all expressions used in the TACO paper~\cite{kjolstad2017tensor}. These real-world applications comprise of algorithms such as factor analysis (e.g. alternating least squares), graph convolutional networks, and tensor factorization and decomposition for domains like machine learning, data analytics, and scientific computing. \Cref{tab:generality-study-taco} lists the sparse tensor algebra features used by each expression and the number of primitives it uses (empty cells denote a primitive is not used). 
We see the primitive counts for level scanners and writers are higher, since they are used for tensor iteration and correspond to the tensor orders of all inputs and outputs respectively. The primitive composition counts also show that most blocks are uniformly used. It is interesting to note that two expressions use all primitive types. 
In addition, we automatically lowered all graphs to our simulator and checked for functional correctness on the set of all real and integer SuiteSparse matrices~\cite{davis2011university} and FROSTT tensors~\cite{frosttdataset} that fit into memory and the Facebook tensor~\cite{viswanath2009evolution}.

\subsection{Ablation Study on the Utility of SAM Blocks}
\label{sec:eval-primitive-study}

\begin{table}[t]
  \caption{
    The number of sparse tensor algebra algorithms from the TACO website that are not expressible if a SAM primitive is removed. The \textbf{All} and \textbf{Unique} columns analyze all input algorithms and distinct algorithms, respectively.
    \label{tab:primitive-study}
  }
  %\footnotesize
  \centering
  \resizebox{\columnwidth}{!}{
  \begin{filecontents*}{figures/primitive-study.csv}
1,Comp. Level Scanner,2773,19363,72.23,81.38
2,Comp. + Uncomp. Level Scanners,3814,23713,99.35,99.66
3,Repeater,3162,19924,82.37,83.74
4,Unioner,600,2229,15.63,9.37
5,Intersecter keep Locator,720,2715,18.75,11.41
6,Intersecter w/ Locator Removed,1878,15777,48.92,66.31
7,Adder,1023,3118,26.65,13.1
8,Multiplier,3220,20986,83.88,88.2
9,Reducer,3008,20036,78.35,84.21
10,Coordinate Dropper,617,2292,16.07,9.63
11,Comp. Level Writer,1075,5525,28,23.22
12,Comp. + Uncomp. Level Writers,3698,23260,96.33,97.76
\end{filecontents*}
\pgfplotstabletypeset[
    col sep=comma,
    numeric type,
    column type={r},
        assume math mode,
    clear infinite,
    %string replace={nan}{---},
    fixed,
    fixed zerofill,
    header=false,
    display columns/0/.style={string type,
                              column name={},
                              column type={l},
                              preproc cell content/.code={}},
    display columns/1/.style={string type,
                              column name={SAM Primitive Removed},
                                                          column type={l},
                              preproc cell content/.code={}},
    display columns/2/.style={string type,
                              column name={Unique},
                              preproc cell content/.code={}},
    display columns/3/.style={string type,
                              column name={All},
                              preproc cell content/.code={}},
    display columns/4/.style={string type,
                              column name={Unique},
                              preproc cell content/.code={}},
    display columns/5/.style={string type,
                              column name={All},
                              preproc cell content/.code={}},
    assign column name/.style={
      /pgfplots/table/column name={\sffamily\textbf{#1}}
    },
    every head row/.style={
    before row={\toprule&&\multicolumn{2}{c}{\sffamily\textbf{Expressions Lost }}&\multicolumn{2}{c}{\sffamily\textbf{Percentage (\%)
    }}\\\cmidrule(lr){3-4}\cmidrule(lr){5-6}}, 
    after row=\midrule},
    every last row/.style={after row=\bottomrule},
    ]{figures/primitive-study.csv}
 }
\end{table}

Each SAM block in \Cref{sec:sam} is essential for expressing the domain of sparse tensor algebra for dataflow. In order to demonstrate the usefulness of each primitive, 
we analyze the entire set of algorithms input by users into the \href{http://tensor-compiler.org/}{TACO website}, provided by the TACO authors, and show which algorithms are not expressible if a given SAM primitive does not exist. We use the TACO website as our dataset since it is representative of real-world sparse tensor algebra computations. From the website, users have successfully compiled 23,794 sparse tensor algebra algorithms to date---of which 3,839 were distinct algorithms (unique combinations of expression and format) and 1,745 were unique solely in expression. 
\Cref{tab:primitive-study} shows that removing any SAM primitive limits the expressible algorithms in the domain of sparse tensor algebra. Most blocks are used for most applications, and, moreover, full algorithmic generality requires all the primitives presented in \Cref{sec:sam}. 

\subsection{Asymptotic Tradeoff Analysis}

\begin{figure}[t!]
        \begin{minipage}[t]{.55\linewidth}
            % \centering
      \includegraphics[scale=0.15]{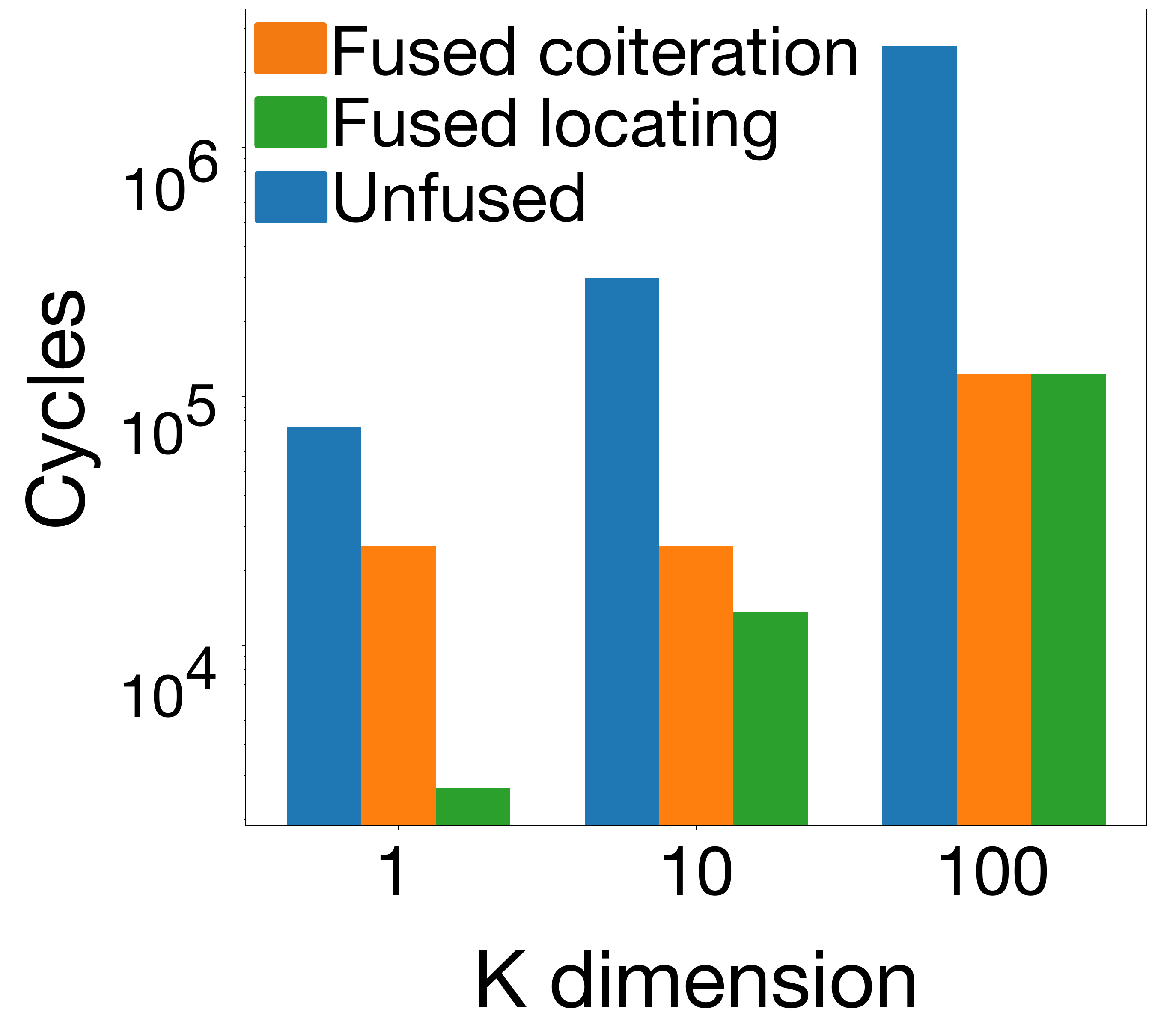}
            \caption{
                Performance of fused and unfused SDDMM algorithms.
                \label{fig:sam-perf-fusion}
            }
        \end{minipage}
        % \hfill{\linewidth}
        \hspace{.01\linewidth}
        \begin{minipage}[t]{.42\linewidth}
            % \centering
             \includegraphics[scale=0.15]{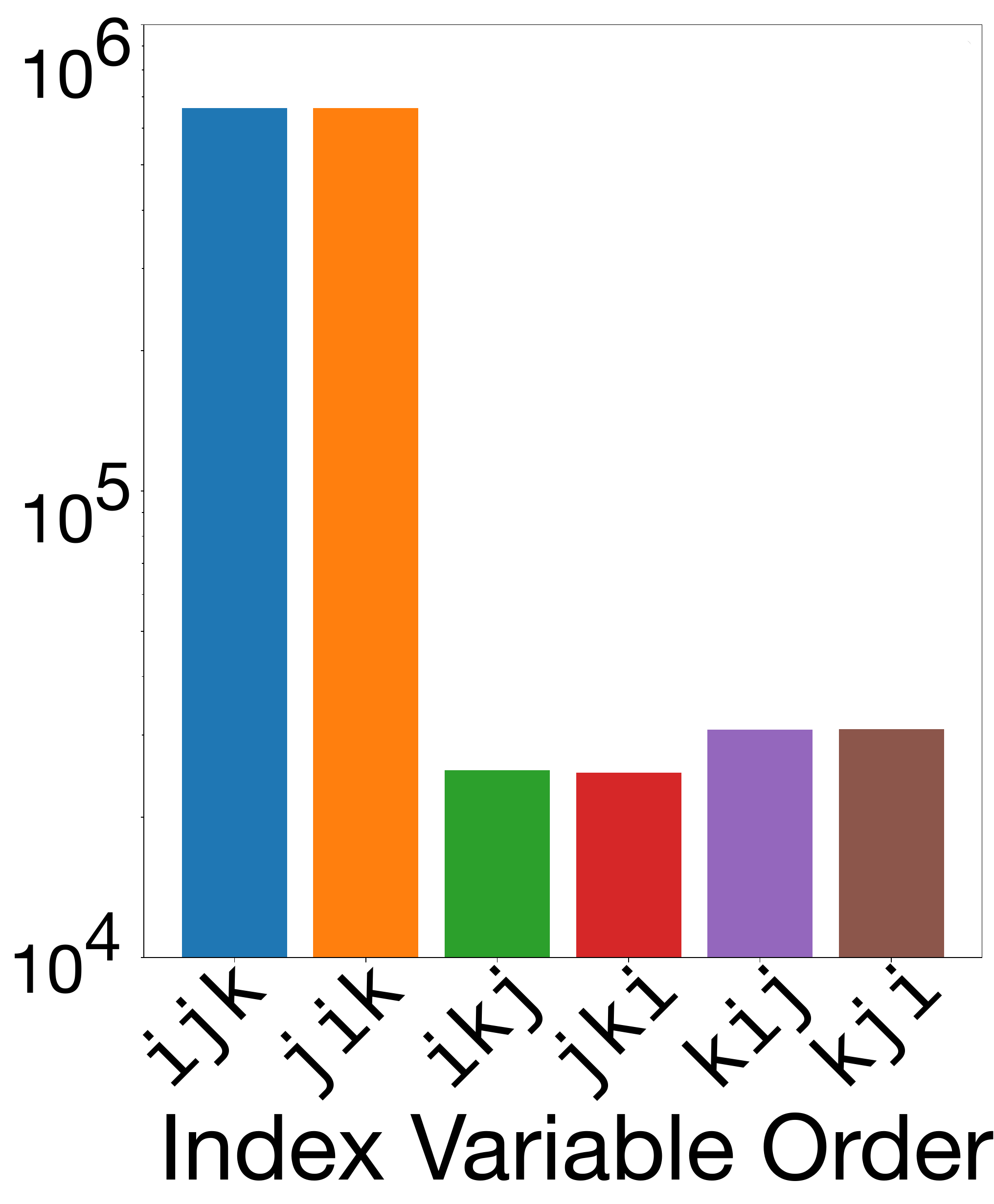}
            \caption{
                Performance of \spmspm{} dataflows. \label{fig:sam-perf-reorder}
            }
        \end{minipage}
        \vspace{-1.5em}
\end{figure}

We next explore the performance attributed to: dataflow ordering, fusion, and various acceleration techniques. While the former fundamentally change the dataflow of the computation, the other optimizations presented are orthogonal and, only affecting a single tensor level, can be used in conjunction with any dataflow.

\subsubsection*{Dataflow Ordering}
\label{sec:eval-reorder}
The index-variable order avoids different data-dependent asymptotic behaviors~\cite{henry2021,yadav2022} and allows for generality in the execution of a particular dataflow \alg{}. We simulate all six permutation orders of $ijk$ for the \spmspm{} expression using two distinct 95\% sparse uniformly random matrices with different dimensions of sizes $I=J=250$ and $K=100$. \Cref{fig:sam-perf-reorder} shows the inner-product algorithms ($ijk$, $jik$) perform the worst for matrix multiply. The linear combination of rows ($ikj$, $jki$) and outer product ($kij$, $kji$) algorithms perform a least an order of magnitude better. The performance is dictated by the order of $k$ since coordinates are filtered out (intersected) at $k$ earlier in the dataflow before repeating along the other dimensions $i, j$. These algorithms differ in their asymptotic complexity~\cite{kjolstad2020sparse,ahrens2022}, so performance differences will increase with increases in sparsity. However, the inner-product algorithm may be more efficient with other data and uses asymptotically less memory for the reduction (a scalar instead of a row). Since the efficiency choice is a tradeoff, sparse hardware should support many processing orders. 

\subsubsection*{Fusion}
\label{sec:eval-fusion}
 
We demonstrate the algorithmic performance advantage of fusion using a common expression from machine learning, the $ijk$-ordered SDDMM $X_{ij} = \sum_k B_{ij}C_{ik}D_{jk}$~\cite{gale2020,bharadwaj2022}.
We generate a $95\%$ sparse uniformly random matrix along with two dense matrices of dimensions $I=J=250$ with a sweep of $K=\{1, 10, 100\}$. 
\Cref{fig:sam-perf-fusion} shows that the unfused implementation performs far worse, since calculating the entire dense matrix multiplication is costly with mostly wasted work. Given the number of nonzeros in $B$ as $\text{nnz}_B$, the unfused computation complexity is proportional to max($\text{nnz}_B*K, \text{locate(nnz}_B)$), while the cost of factorization becomes $I*J*K+\text{locate(nnz}_B)$. The only case where we would want to factorize this expression is when the matrix $B$ is almost fully dense and we have very efficient dense matrix multiplication hardware. But for a sufficiently sparse matrix, a fused expression will perform far better. Efficient sparse hardware must therefore support fused expressions.

We further enhance performance by using locator blocks (\Cref{sec:locate}) to find the sampled $i,j$ values, which is trivial in a dense array.  Interestingly, \Cref{fig:sam-perf-fusion} shows that this advantage becomes negligible as $K$ increases: iteration costs of the dense inner-product dimension $k$ will dominate the computation time, hiding the benefits of locating during intersection. But locating provides significant performance gains when the amount of computation is modest, which is often true in sparse computations.

\subsubsection*{Accelerator Structures}
\label{sec:eval-accel}

\begin{figure*}[ht!]
    \centering
    \includegraphics[width=0.9\linewidth]{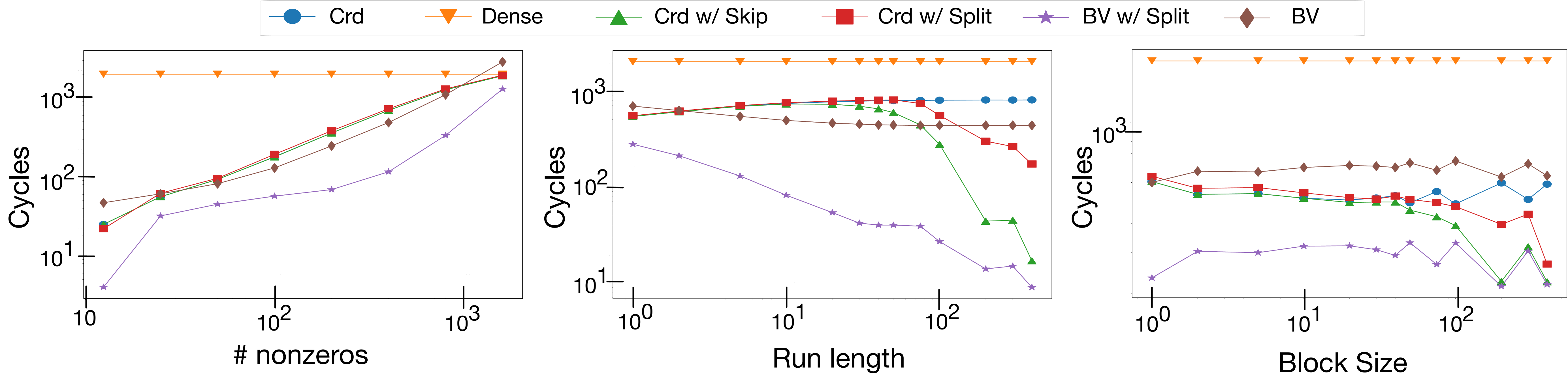} \\[-1.5em]
    {
    \subfloat[Performance vs. sparsity of uniformly random synthetic vectors on a log-log scale]{
        \hspace{.32\linewidth}
        \label{fig:sam-perf-accel-urandom-sf-const}
    }
    \hfill
    \subfloat[Performance vs. run length of synthetic vectors with runs on a log-log scale]{
        \hspace{.29\linewidth}
        \label{fig:sam-perf-accel-runs}
    }
    \hfill
    \subfloat[Performance vs. block size of blocked synthetic vectors on a log-log scale]{
        \hspace{.29\linewidth}
        \label{fig:sam-perf-accel-blocks}
    }}
    \caption{
        Simulated performance of various optimization techniques (compression, splitting, skipping, and bitvectors) for sparse vector sparse vector element-wise multiplication where the vectors have a dense dimension size of 2000.
        \label{fig:sam-perf-all-accel}
    }
\end{figure*}

We next explore different iteration acceleration techniques by comparing various configurations of coordinate-skipping (\Cref{sec:crd-skip}), bitvector iteration (\Cref{sec:bitvectors}), and iteration-splitting (\Cref{sec:tiling}). \Cref{fig:sam-perf-all-accel} compares the performance when both vectors are in the following formats: one uncompressed level (Dense), one compressed coordinate level (Crd), one compressed coordinate level with coordinate-skipping (Crd w/ skip), two compressed coordinate levels (Crd w/ split), one pseudo-dense bitvector level (BV), and two bitvector levels (BV w/ split), also known as a bit-tree. For this set of experiments, we assume the coordinates were already split before this operation\footnote{The splitting operation requires a full scan through the data structure, which for this example is as expensive as the operation itself.} and use the vector-vector element-wise multiply expression $x_i = b_i * c_i$  with both $b$ and $c$ as single dimensional vectors of size $2000$. We use three types of synthetic vectors, namely $urandom$, $runs$, and $blocks$; runs and blocks are shown in \Cref{fig:dataset_runs_blocks}. 
Vectors with $runs$ are pairs of vectors where one vector will have longer stretches of nonzeros between the nonzeros of the other vector. Similarly, $blocks$ are vectors which have dense blocks of nonzeros placed throughout the vector. For both these vectors, the number of nonzeros is 400 (20\%) with the index indicating the size of the runs/blocks in each vector.

\Cref{fig:sam-perf-accel-urandom-sf-const} shows the performance as a function of sparsity for $urandom$ data with bitvector bitwidth $b=64$ and split factor (how many chunks the vector is divided up into) $s=64$, where applicable, and shows the limitations of a single-level bitvector. As the sparsity increases, the compressed coordinate format becomes better than the bitvectors, since bitvectors are still a dense representation. The coordinate-skipping behaves exactly the same as the compressed coordinate format since $urandom$ tensors on average have small (around 1.5) run lengths.

\Cref{fig:sam-perf-accel-runs} shows the utility of coordinate skipping and splitting. As run lengths increase, there are more opportunities to skip invalid input coordinates or avoid computation at the outer-level intersection. The bitvector remains flat since the number of nonzeros remains about the same for various run lengths. This advantage of skipping and splitting remains in the $blocks$ case, without the dependence on block size, since intersections can also be dense. 
Overall, these results show the advantage of the implicit parallelism of bitvectors, but show that they need to be organized hierarchically for robust performance.

\subsection{Modeling Exploration}
\label{sec:eval-model}

\subsubsection*{Stream Analysis}
\label{sec:eval-stream}

We analyze the token breakdown of the \sam{} flattened stream representation and identify that the stream control overhead is modest. We use Custard to compile the \sam{} graph for the matrix identity expression $X_{ij} = B_{ij}$, where $B$ is a sparse DCSR matrix, and count the token types for each coordinate stream at the output of each level scanner. In our simulator, we model streams as Python lists and all control tokens as strings.\footnote{In hardware implementations, however, one possible way to implement the control tokens would be as a tagged-union on the wire. There are alternative implementations, like hardcoding the control token level for each primitive, which removes the need for a stop level in the stream but complicates and hardens the state-machine logic of each primitive.}
We run the expression on 15 matrices of various sizes from the SuiteSparse matrix collection~\cite{davis2011university} (see \Cref{tab:ss-data} in the Appendix for matrix characteristics and selection criteria). 

\begin{figure}[t!]
    \centering
    \includegraphics[width=\columnwidth]{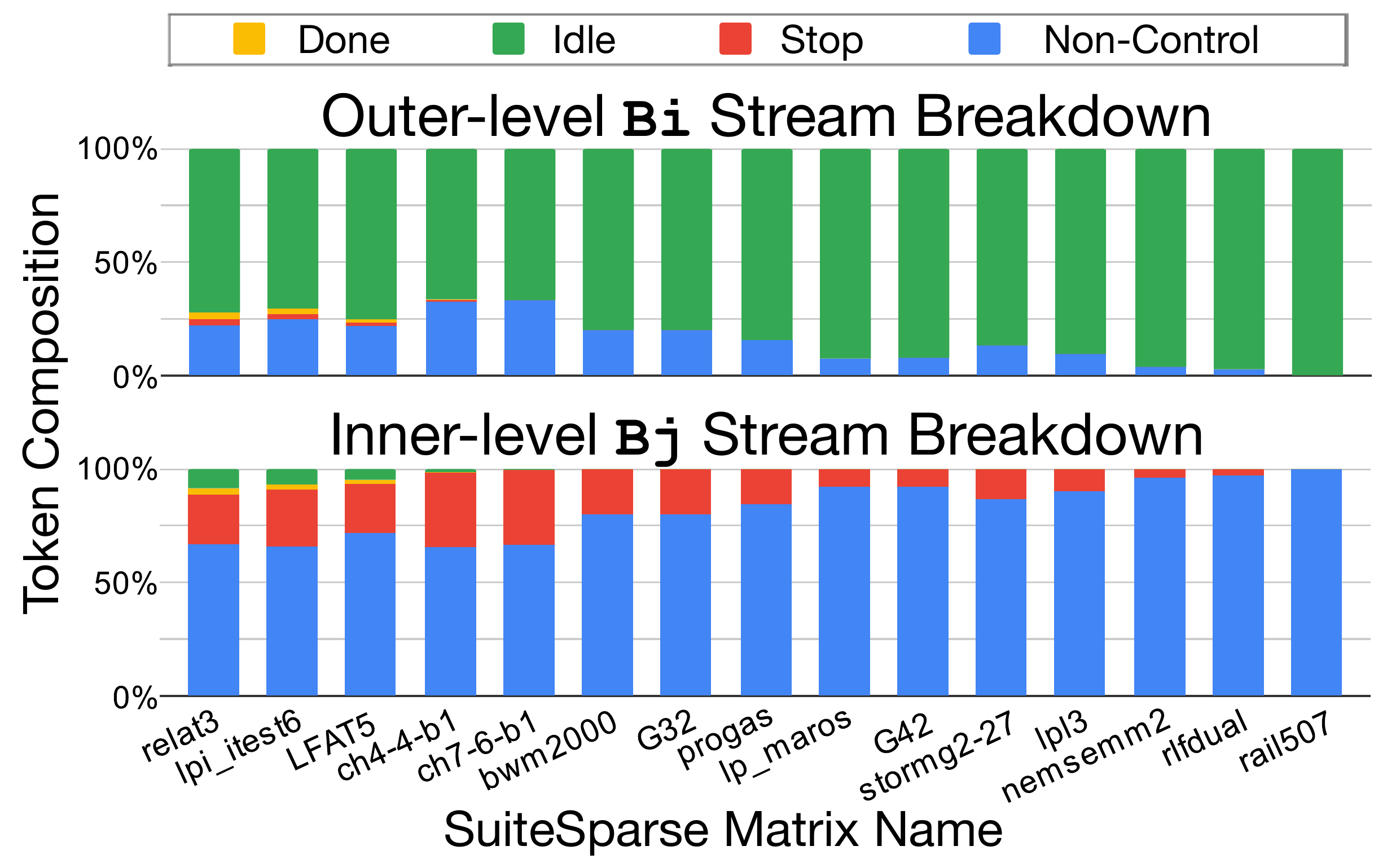}
    \vspace{-4mm}
    \caption{Breakdown of the outer $B_i$ level stream and inner $B_j$ level stream by token type for the matrix identity expression $X_{ij}=B_{ij}$ where $B$ is a sparse DCSR matrix}
    \label{fig:eval-stream-overhead}
\end{figure}

The control token overhead of our representation is reasonable, with an average non-idle control overhead reaching 0.95\% for outer levels and 16.20\% for inner levels as shown in \Cref{fig:eval-stream-overhead}. (\Cref{sec:point-streams} shows the control overhead of the alternative of using non-flattened point streams would be higher.) The average inner-level percentage means that rows have an average of 5 nonzeros, an appropriate number of coordinates for this set of matrices. The outer-level $B_i$ stream and inner-level $B_j$ stream refer to the coordinate stream outputs of the first $B_i$ level scanner and the second $B_j$ level scanner, respectively. We do not show the $B_{vals}$ breakdown since it is the same as $B_j$. Most tokens, on average 83.32\%, on the $B_i$ stream are idle since the $B_i$ level scanner is in the done state while the inner-level iterates through its coordinates. This behavior occurs in compressed arrays, as in \Cref{fig:data-model-memory}, because there are exponentially more coordinates for each lower level of a tensor. The done state of the primitive is efficient as it is idle and avoids computation activity. Improving efficiency and utilization of idle primitives could include switching the outer level scanner to other tasks through time multiplexing, which we leave as future work.
At the lower level of the matrix, control overhead is dominated by stop tokens. The stop token overhead ranges from 0.12\% (for \texttt{rail507}) to 33.26\% (for \texttt{ch7-6-b1}). Again, these breakdowns are reasonable since higher percentages of stop tokens occur only in small matrices.

\subsubsection*{Modeling Hardware with Finite Constraints}
\begin{figure}[t!]
    \centering
    \includegraphics[width=0.9\linewidth]{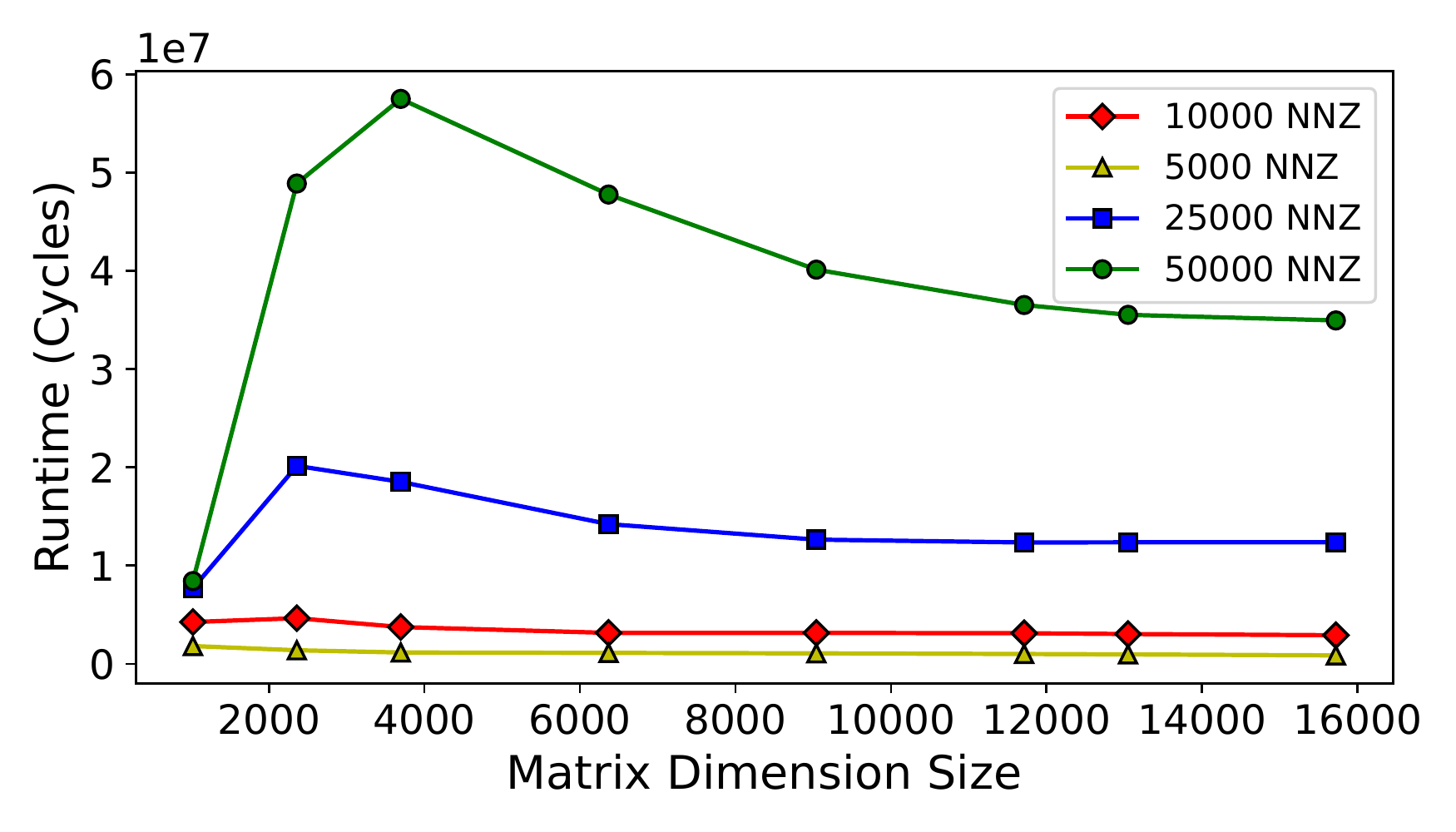}
    \vspace{-4mm}
    \caption{Recreation of ExTensor's ``\spmspm{} performance across varying dimension sizes with a constant number of nonzeros per matrix'' study using our \sam{} simulator.
    \label{fig:extensor-mem-model}}
\end{figure}

Although \sam{} is an abstract machine with infinite resources, it can also represent finite hardware with finite memory. ExTensor~\cite{hegde2019extensor} is one design point in the space of sparse tensor algebra accelerators, and \sam{} is sufficiently expressive to model it. We find that \sam{} can recreate the performance characteristics of ExTensor's evaluation. We recreate the synthetic data study, Figure 19 Section 8.4, in the ExTensor paper that measures ``\spmspm{} performance across varying dimension sizes with a constant number of nonzeros per matrix'' as shown in~\Cref{fig:extensor-mem-model}. Our \sam{} model contains the hierarchical coordinate skipping, fixed-memory tiling, sparse tile skipping, and n-buffering techniques of ExTensor. Our performance matches the three performance regions of the ExTensor study: increasing runtime due to more non-empty tiles at small dimensions, decreasing runtime due to sparse tile skipping, and decreasing runtime with saturating performance. Concretely, we model two levels of memory hierarchy, a last-level buffer (LLB) and a processing element buffer (PEB). \sam{} is used to sequence the sparse tile coordinates including the CPU loop-nests and using a DRAM bandwidth of 68.256 GB/s, an LLB size of 17MB, and a processing element (PE) tile size of 128$\times$128 (configured using implementation-specific information).  

\subsection{Backend Case Studies}
\label{sec:classification}

\begin{figure*}[ht!]
    \centering
    \includegraphics[width=\linewidth]{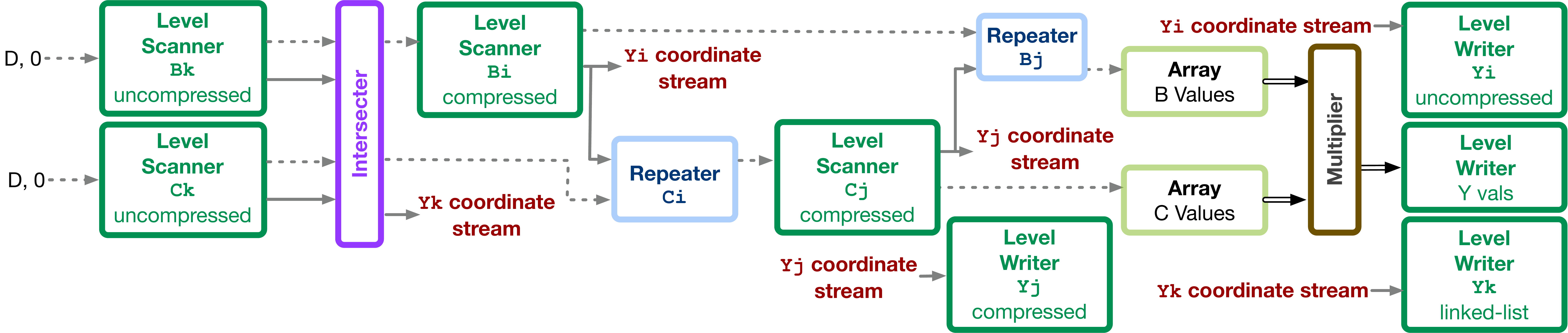}
    \caption{The SAM dataflow graph for \spmspm{} that represents the OuterSPACE multiply phase, which is followed by the OuterSPACE merge phase. Compare this outer product graph to the linear combination graph in \Cref{fig:sam-gemm}.
    }
    \label{fig:outerspace-mult}
\end{figure*}

By construction, we designed \sam{} to easily represent dataflow hardware. To evaluate its likeness and ability to bind to hardware, we qualitatively analyze how \sam{} is able to represent fixed-function and reconfigurable dataflow backends including Gamma \cite{zhang2021gamma}, OuterSPACE~\cite{pal2018outerspace}, 
ExTensor~\cite{hegde2019extensor}, and Capstan~\cite{rucker2021capstan}. For example, Gamma's dataflow is similar to \Cref{fig:sam-gemm}. The main difference is that Gamma adds a parallelizer after the intersection unit and then uses a multi-input vector reducer to rejoin the parallel threads.

For space reasons, we only provide a concrete \sam{} graph for OuterSPACE (see \Cref{fig:outerspace-mult}), which leverages an outer-product dataflow ($k \rightarrow{}i \rightarrow{} j$). We chose OuterSPACE because it factorizes \spmspm{} into two stages: a multiply phase ($Y_{ikj} = B_{ik} C_{kj}$) and a merge phase ($X_{ij} = Y_{ikj}$), thus showing how \sam{} supports factorization. For efficiency, $B_{ik}$ and $C_{kj}$ are respectively stored in column-major and row-major order. The first phase computes outer products between all columns of $B$ and all rows of $C$ and stores the partial result into a 3-dimensional tensor $Y_{ikj},$ as shown in \Cref{fig:outerspace-mult}. To efficiently merge  in the next phase, the intermediate result $Y$ is stored in $ikj$-order, which is discordant with the dataflow $kij$. To efficiently support a discordant write of the tensor streams, OuterSPACE utilizes a linked-list representation as the level-format for $k$. Because our \wrscan{}  is not restricted to a specific representation, \sam{} supports this dataflow. The merge phase (not shown to conserve space) then accumulates the partial product $Y_{ikj}$ from the previous phase into a final result $X_{ij}$. This dataflow consists of three cascaded \rdscan{s} to generate the values $Y_{ikj}$ that need to be summed, a vector reducer to sum the $k$ dimension, and three \wrscan{s} to store the $X_{ij}$ results. 

ExTensor's Stream Coordinator~\cite{hegde2019extensor} is naturally representable with \sam{}. It is a hardware unit consisting of, in order, their Stream Sequencer, two of their Scanners, their Intersection unit, and two Data Storage units. The Stream Sequencer (and its \texttt{Configure()} command) is representable with two \sam{} repeater primitives, with the added benefit that \sam{'s} repeaters do not have to be pre-configured. We represent their Scanner as a composition of $n$ \sam{} level scanners with coordinate-skipping since their Scanner can scan through $n$ levels of a fibertree. Finally, their Intersect and Data Storage units are equivalent to our intersecter and array primitives. We note that each ExTensor Coordinator is fixed for two input tensors (hence two Scanners with Data Storage) with the intersection always occurring at the last level of each unit. These implementation choices limit Extensor-like \sam{} graphs to a subset of the entire space producible by \sam{}. 

Finally, we analyze the Capstan~\cite{rucker2021capstan} specialized loop-header hardware, since it is one of the main contributions of that work. Capstan can represent this specialized hardware as a two-operand bitvector scanner that can be configured with \texttt{or} or \texttt{and}. Using \sam{}, the hardware is equivalent to two bitvector level scanners (one for each tensor) followed by an intersecter for \texttt{and} or a unioner for \texttt{or}. The output values produced by the Capstan hardware: addresses for both tensors, a store address, and a dense index correspond to \sam{'s} post-intersect/union reference streams, result level writer address generation, and post-intersect/union coordinate stream, respectively. The vectorized loop bodies in Capstan are representable using n-lane stream buses (bundles), \sam{} arrays to get the data, a single \sam{} ALU, and level writers. Again, the class of \sam{} graphs that represent a single Capstan loop-header is fixed to two sparse operands with only a subset of \sam{'s} expressibility. Additionally, the Capstan-like \sam{} only iterates through bitvectors, which is great for vectorization but is fixed to a pseudo-dense iteration space.

\section{Conclusion}
We introduced the \samlong{}, an abstract machine model for both reconfigurable and fixed-function spatial dataflow accelerators for sparse tensor algebra. Our design led to a stream representation and a small set of physical blocks with well-defined interfaces. Our Custard compiler demonstrates \sam{}'s utility as a compiler target. In addition, the flexibility and generality of \sam{} let us fairly evaluate optimization and dataflow alternatives for accelerating sparse tensor algebra algorithms. We hope that the Sparse Abstract Machine model will enable the microarchitectural design of future accelerators and inform the design decisions of architects. We also hope that compiler designs like Custard, targeting an abstract machine for portability, will improve the programmability and usability of this space.

\begin{acks}
We thank Manya Bansal, Zimren Dixon, Scott Kovach, Zachary Myers, Aviral Pandey, Alexander Rucker, Matthew Sotoudeh, Shiv Sundram, Joseph Tan, and Rohan Yadav for their helpful feedback. We would also like to thank Ajay Brahmakshatriya, Jake Ke, Kalhan Koul, Qiaoyi Li, Keyi Zhang, Saman Amarasinghe, Riyadh Baghdadi, and Priyanka Raina for discussion and help with evaluation. Olivia Hsu was supported by an NSF GRFP Fellowship, Maxwell Strange was supported by the Apple Stanford Electrical Engineering PhD Fellowship in Integrated Systems, and Ritvik Sharma was supported by the Stanford Graduate Fellowship. This work was supported in part by the NSF under grant numbers 1937301, 2028602, CCF-1563078, and 1563113; by DARPA under the Domain-Specific System on Chip (DSSoC) Program; and by the DoE National Nuclear Security Administration (NNSA) under grant number DE-NA0003965.  
This research was also supported in part by the Stanford Agile Hardware (AHA) Center, Google Research Scholar program, and Stanford Data Analytics for What’s Next (DAWN) Affiliate Program. Any opinions,
findings, and conclusions or recommendations expressed in this material are those of the authors
and do not necessarily reflect the views of the aforementioned funding agencies.
\end{acks}

\appendix
% LaTeX template for Artifact Evaluation V20201122
%
% Prepared by 
% * Grigori Fursin (cTuning foundation, France) 2014-2020
% * Bruce Childers (University of Pittsburgh, USA) 2014
%
% See examples of this Artifact Appendix in
%  * SC'17 paper: https://dl.acm.org/citation.cfm?id=3126948
%  * CGO'17 paper: https://www.cl.cam.ac.uk/~sa614/papers/Software-Prefetching-CGO2017.pdf
%  * ACM ReQuEST-ASPLOS'18 paper: https://dl.acm.org/citation.cfm?doid=3229762.3229763
%
% (C)opyright 2014-2020
%
% CC BY 4.0 license
%

% \documentclass[sigconf]{acmart}

% \usepackage{hyperref}

% \begin{document}

% \special{papersize=8.5in,11in}

%%%%%%%%%%%%%%%%%%%%%%%%%%%%%%%%%%%%%%%%%%%%%%%%%%%%
% When adding this appendix to your paper, 
% please remove above part
%%%%%%%%%%%%%%%%%%%%%%%%%%%%%%%%%%%%%%%%%%%%%%%%%%%%

\section{Artifact Appendix}

%%%%%%%%%%%%%%%%%%%%%%%%%%%%%%%%%%%%%%%%%%%%%%%%%%%%%%%%%%%%%%%%%%%%%
\subsection{Abstract}

This appendix describes how to set up and run our Sparse Abstract Machine (SAM) Python simulator and the C++ CUSTARD compiler, which compiles from concrete index notation (CIN) to SAM graphs (represented and stored in the DOT file format). The appendix also describes how to reproduce the quantitative experimental results in this paper. The artifact can be executed with any X86-64 or M-series Apple machine with Docker, Python 3, Git, and Bash support, at least 32 GB of RAM, and more than 20 GB of disk space.

\subsection{Artifact Check-List (Meta-Information)}

% {\em Obligatory. Use just a few informal keywords in all fields applicable to your artifacts
% and remove the rest. This information is needed to find appropriate reviewers and gradually 
% unify artifact meta information in Digital Libraries.}

{\small
\begin{itemize}
  % \item {\bf Algorithm: }
  % \item {\bf Program: }
  
  \item {\bf Compilation: }
  C++ compiler (either \texttt{gcc} or \texttt{clang}). The \texttt{gcc} 9.4.0 compiler is included with the Docker image. A fork of the TACO compiler (found \href{https://github.com/weiya711/taco/tree/cf8f007bab940527c48fb0479236b3928b32c1b0}{here}) is included as a submodule in our artifact (fork \texttt{weiya711/taco}, commit hash \texttt{cf8f007}). 
  % \item {\bf Transformations: }
  % \item {\bf Binary: } The compiler binaries are included in the docker
  % \item {\bf Model: }
  
  \item {\bf Data set: }
  Suitesparse Matrix Market matrices (a script to download the dataset is included and the full dataset can be found at \href{https://sparse.tamu.edu/}{https://sparse.tamu.edu/}), Frostt Tensor Dataset tensors (a script to download the dataset is included and the full dataset can be found at \href{http://frostt.io/}{http://frostt.io/}), and synthetically generated matrices/higher-order tensors (included).
  
  \item {\bf Run-time environment: }
  Docker, git, Python 3, and bash need to be installed on the local machine. Docker is available for many operating systems. Proficiency in bash and git is recommended.
  
  \item {\bf Hardware: }
  Any conventional x86 CPU with at least 32 GB of RAM should work.
  % \item {\bf Run-time state: }
  % \item {\bf Execution: }
  
  \item {\bf Metrics: }
  Cycles (modeled as iteration counts in our simulator and source code), expression counts, and primitive counts
  
  \item {\bf Output: }
  Terminal outputs, files, tables, and graphs (PDF figures, PNG figures, and DOT file format~\cite{ellson2002graphviz} graphs). Expected results are included in the submitted paper. 

  \item {\bf Experiments: }
  All steps are detailed in the \texttt{README.md} in \\ 
  \href{https://github.com/weiya711/sam-artifact}{https://github.com/weiya711/sam-artifact}. The steps include pulling a Docker image and running/attaching a container, running scripts within the docker, running one Python 3 script locally outside of the Docker to copy results, and verifying result images/files. The experiments should have less than 5\% variation since the simulator is deterministic. The 5\% variation is caused by different data patterns in synthetic data generation (even with sparsity held constant due to random statistics). However, these variations do not affect the paper's conclusions.
  
  \item {\bf How much disk space required (approximately)?: } 
  Approximately 20GB of space should be sufficient.
  
  \item {\bf How much time is needed to prepare the workflow (approximately)?: }
  About 10-15 minutes.
  
  \item {\bf How much time is needed to complete experiments (approximately)?: }
  To complete all experiments it takes approximately 65 hours. We also include scripts to complete a subset of the experiments, which include~\Cref{tab:generality-study-taco}, \Cref{tab:primitive-study}, \Cref{fig:sam-perf-fusion}, 
 \Cref{fig:sam-perf-reorder}, \Cref{fig:sam-perf-all-accel}, \Cref{fig:eval-stream-overhead}, and only 8 points in ~\Cref{fig:extensor-mem-model}, that takes about 10 hours to run on a standard machine.
  
  \item {\bf Publicly available?: }
  Yes, on Github at the \textit{sam} repository \\
  (\href{https://github.com/weiya711/sam}{https://github.com/weiya711/sam}) for active development of source code
  and at the \textit{sam-artifact} repository \\
  (\href{https://github.com/weiya711/sam-artifact}{https://github.com/weiya711/sam-artifact}) for the artifact evaluation of this paper. The specific commits for this artifact are tagged with \texttt{asplos23-ae} in both repositories. 
  
  \item {\bf Code licenses (if publicly available)?: }
  MIT License
 
  \item {\bf Workflow framework used?: }
  Docker
  
  \item {\bf Archived (provide DOI)?: }
  Yes, the DOI is \\
  \href{https://doi.org/10.5281/zenodo.7591742}{https://doi.org/10.5281/zenodo.7591742}~\cite{hsu2023sam-artifact}.

\end{itemize}
}

%%%%%%%%%%%%%%%%%%%%%%%%%%%%%%%%%%%%%%%%%%%%%%%%%%%%%%%%%%%%%%%%%%%%%
\subsection{Description}

\subsubsection{How to Access}

The code repository for this submission can be downloaded from \href{https://github.com/weiya711/sam-artifact}{https://github.com/weiya711/sam-artifact}. The repository includes a Dockerfile from which a Docker image can be built for full evaluation of the artifact.

\subsubsection{Hardware dependencies}

We recommend a machine with a conventional x86 CPU and at least 32GB of memory. We found that some of the experiments will be OOM killed on a machine with only 16GB of memory.

\subsubsection{Software Dependencies}

Evaluation of the artifact requires a machine with Docker  and Python 3 installed. We tested the artifact evaluation on the following configurations and found them to work: Ubuntu 20.04/Docker 20.10.12/Python 3.8 (AMD-based machine), and MacOS 13.1/Docker 20.10.22/Python 3.9 (Intel-based machine). We expect other versions of MacOS, Ubuntu, Docker, and Python 3 configurations to work as well.

\subsubsection{Data sets}
\label{sec:appendix-datasets}

\begin{figure}[htbp]
    \centering
    % \hfill
    % \begin{minipage}[t]{0.4\linewidth}
        \includegraphics[width=\columnwidth]{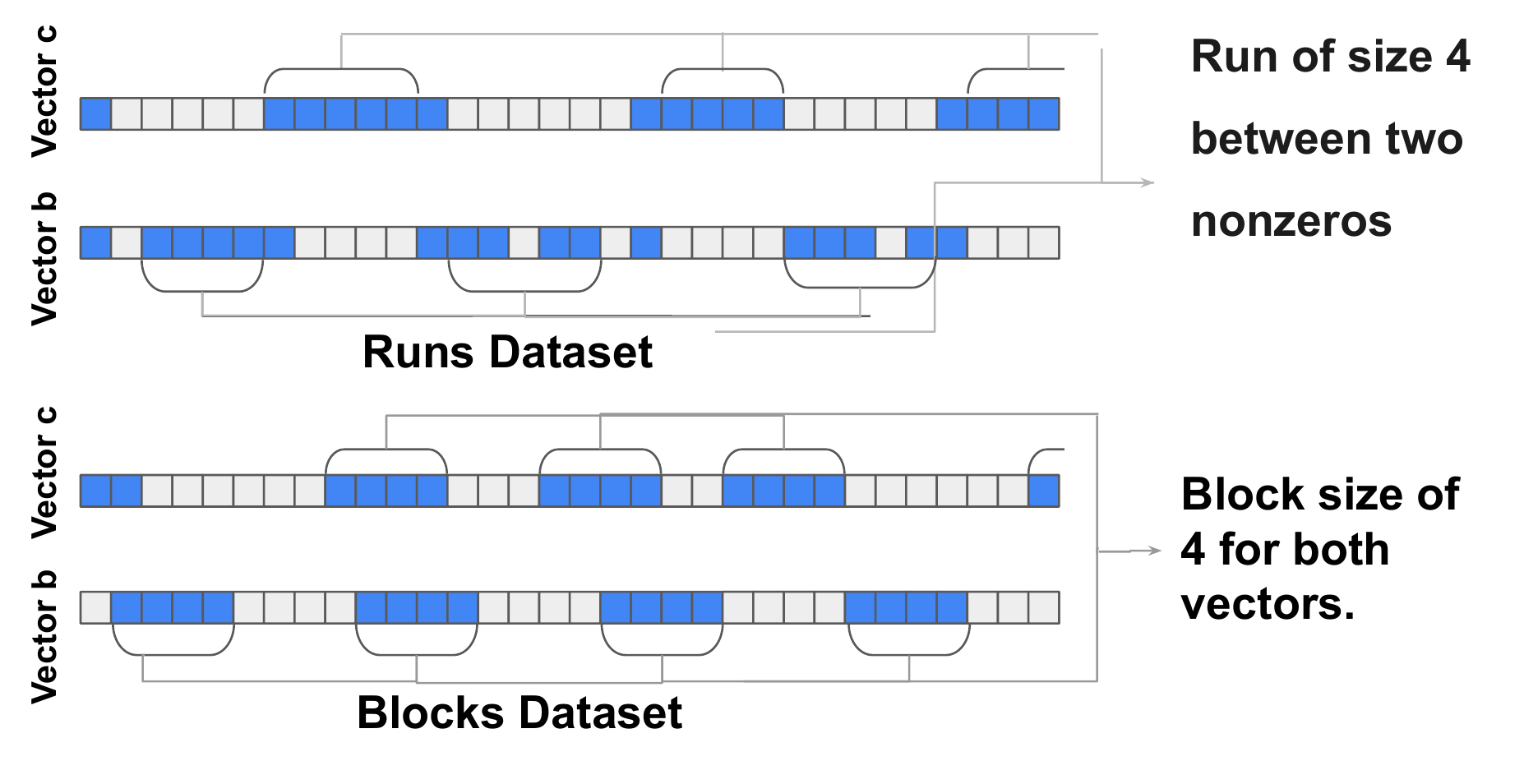}
        \caption{
            Sample $runs$ and $blocks$ vector data patterns used for our synthetic data generation in \Cref{sec:eval-accel}.
            \label{fig:dataset_runs_blocks}
        }
    % \end{minipage}
\end{figure}

\begin{table}[htbp]
  \caption{
    Matrices from the SuiteSparse matrix collection~\cite{davis2011university} used to analyze the overhead of our stream representation in matrix identity (\Cref{sec:eval-stream}). We randomly selected each set of 5 matrices (delineated in the table above) from the smallest, median, and largest 50 SuiteSparse matrices---based on dense dimension size---that would fit in memory. 
    \label{tab:ss-data}
  }
  \centering
  %\footnotesize
  \scriptsize
  \setlength{\tabcolsep}{2pt}
  %\tabulinesep=0pt
  \begin{tabu}{ll|c|rr}
    \toprule
    \rowfont{\sffamily\bfseries} 
    Name & Domain & Dimensions & Nonzeros & Density (\%) \\
    \midrule
    relat3 & Combinatorics & $8 \times 5$ & 24 & 60.0 \\
    lpi\_itest6 & Linear Programming & $11 \times 17$ & 29 & 15.5 \\
    LFAT5 & Model Reduction & $14 \times 14$ & 46 & 23.5 \\
    ch4-4-b1 & Combinatorics & $72 \times 16$ & 144 & 12.5 \\
    ch7-6-b1 & Combinatorics & $630 \times 42$  & 1260 & 4.8 \\
    \midrule
    bwm2000 & Chemical Process Simulation & $2000 \times 2000$ & 7996 & 0.2 \\
    G32 & Undirected Weighted Random Graph & $2000 \times 2000$ & 8000 & 0.2 \\
    progas & Linear Programming & $1650 \times 1900$ & 8897 & 0.3 \\
    lp\_maros & Linear Programming & $846 \times 1966$ & 10137 & 0.6 \\
    G42 & Undirected Weighted Random Graph & $2000 \times 2000$ & 23558 & 0.6 \\
    \midrule
    stormg2-27 & Linear Programming &  $14,439 \times 37,485$  & 94274 & 0.02 \\
    lpl3 & Linear Programming & $10,828 \times 33,686$  & 100525 & 0.03 \\
    nemsemm2 & Linear Programming & $6943 \times 48,878$ & 182012 & 0.05 \\
    rlfdual & Linear Programming & $8052 \times 74,970$ & 282031 & 0.05 \\
    rail507 & Linear Programming & $507 \times 63,516$ & 409856 & 1.3 \\
    \bottomrule
  \end{tabu}%\vfill
\end{table}
The evaluation requires matrices from the Suitesparse Matrix Market dataset (script to download the dataset is included, full dataset can be found at \href{https://sparse.tamu.edu/}{https://sparse.tamu.edu/}), the Frostt Tensor Dataset (script to download the dataset is included, full dataset can be found \href{http://frostt.io/}{http://frostt.io/}), and synthetically generated matrices/higher-order tensors (included in the artifact evaluation). The synthetic data generation pattern for~\Cref{sec:eval-accel} is shown in~\Cref{fig:dataset_runs_blocks}.

% \subsubsection{Models}

%%%%%%%%%%%%%%%%%%%%%%%%%%%%%%%%%%%%%%%%%%%%%%%%%%%%%%%%%%%%%%%%%%%%%
\subsection{Installation}

To install, first clone the \href{https://github.com/weiya711/sam-artifact}{\textit{sam-artifact}} repository to the local machine and initialize all submodules, then build the docker image:
\begin{minted}{bash}
$ git clone https://github.com/weiya711/sam-artifact
$ cd sam-artifact
$ git submodule update --init --recursive
$ docker build -t sam-artifact .
\end{minted}
The docker container can be started with the following command:
\begin{minted}{bash}
$ docker run -d -it --rm sam-artifact bash
\end{minted}
A docker container ID will be printed upon completion of this command, and the container can be attached to with:
\begin{minted}{bash}
$ docker attach <CONTAINER_ID>
\end{minted}
Once inside the container, a fire test can be conducted via the commands:
\begin{minted}{bash}
$ cd /sam-artifact/sam/
$ python scripts/collect_node_counts.py
\end{minted}
%%%%%%%%%%%%%%%%%%%%%%%%%%%%%%%%%%%%%%%%%%%%%%%%%%%%%%%%%%%%%%%%%%%%%
\subsection{Experimental Workflow}

The experimental workflow for this artifact includes running a set of scripts within the Docker environment to generate tables/figures from the paper. The complete instructions can be found in the \texttt{README.md} included within the \href{https://github.com/weiya711/sam-artifact}{\textit{sam-artifact}} repository.

%%%%%%%%%%%%%%%%%%%%%%%%%%%%%%%%%%%%%%%%%%%%%%%%%%%%%%%%%%%%%%%%%%%%%
\subsection{Evaluation and Expected Results}

The following subsection includes information on how to reproduce all the automatically generated result figures/tables in the paper. We note that the left-hand side of Table 1 (Sparse Tensor Algebra Features) and Figure 16 were manually derived by the authors and have no associated source code.

Tables 1, Table 2 and Figures 11--14 can be generated with the following commands:
\begin{minted}{bash}
# In Docker Container
$ cd /sam-artifact
$ source scripts/generate_all_results.sh
ctrl-p ctrl-q   # Detach from Docker container
# In local machine
$ python sam/scripts/artifact_docker_copy.py \ 
    --output_dir <OUTPUT_DIRECTORY> \
    --docker_id <DOCKER_ID>
\end{minted}

The expected results for the above commands are:
\begin{itemize}
    \item \textbf{Table 1}: The standard output from the fire test, which is also saved at \texttt{/sam-artifact/sam/tab1.log} in the Docker container, should match the right hand side of~\Cref{tab:generality-study-taco}. The left-hand side (Sparse Tensor Algebra Features) contains manually derived summaries of the expressions.
    \item \textbf{Table 2}: The file \texttt{/sam-artifact/taco-website/tab2.log} in the
    Docker container should match~\Cref{tab:primitive-study}.
    \item \textbf{Figure 11}: The file \texttt{fig11.pdf} on the local machine should match~\Cref{fig:sam-perf-fusion}.
    \item \textbf{Figure 12}: The file \texttt{fig12.pdf} on the local machine should match~\Cref{fig:sam-perf-reorder}.
    \item \textbf{Figure 13}: The files \texttt{fig13a.pdf}, \texttt{fig13b.pdf}, and \\
    \texttt{fig13c.pdf} on the local machine should match~\Cref{fig:sam-perf-accel-urandom-sf-const}, \Cref{fig:sam-perf-accel-runs}, and \Cref{fig:sam-perf-accel-blocks} respectively.
    \item \textbf{Figure 14}: The file \texttt{fig14.pdf} on the local machine should match~\Cref{fig:eval-stream-overhead}.
\end{itemize}

Figure 15 generation is time-consuming to compute so we have given three options---one data point, a few data points, or all data points---each with size configurations, leading to 5 options total.

\begin{minted}{bash}
# In Docker container
$ cd /sam-artifact/sam
# [Option 1] Choose and run one point from Figure 15
$ ./scripts/single_point_memory_model_runner.sh \
    extensor_<NNZ>_<DIMSIZE>.mtx
# [Option 2] Run eight points from Figure 15
$ ./scripts/few_points_memory_model_runner.sh <GOLD>
# [Option 3] Run all points from Figure 15
$ ./scripts/full_memory_model_runner.sh <GOLD>
ctrl-p ctrl-q   # Detach from Docker container
# In local machine
$ python sam/scripts/artifact_docker_copy.py \ 
    --output_dir <OUTPUT_DIRECTORY> \
    --docker_id <DOCKER_ID>
\end{minted}

where \texttt{NNZ $\in$ \{5000, 10000, 25000, 50000\}},
\texttt{DIMSIZE $\in$ range(1024,15721,1336)}, and \texttt{GOLD $\in$ \{0,1\}},
where 0 means no gold checking and 1 includes gold checking. 
A single point run has gold enabled by default and runs for between 20 minutes to 17 hours, depending on which \texttt{NNZ} and \texttt{DIMSIZE} combination is chosen. The few points script takes approximately 8 hours to run with no gold and 19 hours to run with gold enabled. The full script takes approximately 64 hours to run with no gold and 92 hours to run with gold enabled.

\begin{itemize}
    \item \textbf{Figure 15}: The file \texttt{fig15.pdf} on the local machine should contain a subset of scatter points that match~\Cref{fig:extensor-mem-model}.
    \item \textbf{Figure 16:} This figure is manually derived.
\end{itemize}
%%%%%%%%%%%%%%%%%%%%%%%%%%%%%%%%%%%%%%%%%%%%%%%%%%%%%%%%%%%%%%%%%%%%%
\subsection{Experiment Customization}

Detailed experiment customization can be found in the \textit{sam-artifact} \texttt{README.md} section titled \href{https://github.com/weiya711/sam-artifact#Optional-How-To-Reuse-Artifact-Beyond-the-Paper}{\textbf{How to Reuse Artifact Beyond the Paper}}.

%%%%%%%%%%%%%%%%%%%%%%%%%%%%%%%%%%%%%%%%%%%%%%%%%%%%%%%%%%%%%%%%%%%%%
% \subsection{Notes}

%%%%%%%%%%%%%%%%%%%%%%%%%%%%%%%%%%%%%%%%%%%%%%%%%%%%
% When adding this appendix to your paper, 
% please remove below part
%%%%%%%%%%%%%%%%%%%%%%%%%%%%%%%%%%%%%%%%%%%%%%%%%%%%

% \end{document}

% \clearpage

% \onecolumn
% \begin{multicols}{2}
%    \bibliographystyle{ACM-Reference-Format}
%    \bibliography{references}
% \end{multicols}

%% Add \addtolength{\textheight}{-5.5cm} before last page to balance


\end{document}